\documentclass[10pt,twocolumn,twoside]{IEEEtran}

\usepackage{graphicx} 
\usepackage{todonotes}
\usepackage{amsmath, amssymb}
\usepackage{verbatim}
\usepackage[nolist]{acronym}
\usepackage{subcaption}
\usepackage{algorithm}
\usepackage{algorithmic}
\usepackage{booktabs, multirow, colortbl}
\usepackage{cite}

\title{A Perception vs. Distortion Perspective on Score-Based Generative Channel Estimation}

\author{Marco~Skocaj*,\thanks{*Marco Skocaj and Lukas Eller contributed equally to this work.}~Lukas~Eller*,~Mate~Boban\thanks{All authors are with the Advanced Wireless Technology Laboratory, Huawei Heisenberg Research Centre, Munich, Germany.\\Emails: {\emph{name}.\emph{surname}@huawei.com}}
}
\date{December 2025}

\begin{document}

\begin{acronym}
    \acro{AI}{Artificial Intelligence}
    \acro{AWGN}{Additive White Gaussian Noise}
    \acro{BS}{Base Station}
    \acro{CDM}{Conditional Diffusion Model}
    \acro{CFG}[CFG]{classifier free guidance}
    \acro{CME}[CME]{Conditional Mean Estimator}
    \acro{CSI}{Channel State Information}
    \acro{CSIT}{CSI at the transmitter}
    \acro{DDPM}[DDPM]{Denoising Diffusion Probabilistic Model}
    \acro{DL}{Downlink}
    \acro{DM}{Diffusion Model}
    \acro{E2E}{End-to-End}
    \acro{ERM}[ERM]{Empirical Risk Minimization}
    \acro{GAN}[GAN]{Generative Adversarial Network}
    \acro{GMM}{Gaussian mixture model}
    \acro{GCS}{Generalized Cosine Similarity}
    \acro{LS}{Least Squares}
    \acro{MIMO}{Multiple-Input-Multiple-Output}
    \acro{MI}{mutual information}
    \acro{ML}{machine learning}
    \acro{MMSE}[MMSE]{minimum mean squared error}
    \acro{MSE}[MSE]{mean squared error}
    \acro{NMSE}{normalized mean squared error}
    \acro{NR}{New Radio}
    \acro{OFDM}{Orthogonal Frequency Division Multiplexing}
    \acro{PRB}{Physical Resource Block}
    \acro{SAA}{sample average approximation}
    \acro{SDE}{stochastic differential equation}
    \acro{SE}{spectral efficiency}
    \acro{SGM}{score-based generative model}
    \acro{SGCS}{Squared Generalized Cosine Similarity}
    \acro{SL}{Supervised Learning}
    \acro{SMLD}[SMLD]{score matching with Langevin dynamics}
    \acro{SNR}{signal-to-noise ratio}
    \acro{SRS}{Sounding Reference Signal}
    \acro{SU}{Single-User}
    \acro{SVD}{Singular Value Decomposition}
    \acro{TDD}{Time Division Duplex}
    \acro{UE}{User Equipment}
    \acro{UL}{Uplink}
    \acro{VAE}{variational autoencoder}
    \acro{SBGM}{Score-based Generative Model}
\end{acronym}

\maketitle

\begin{abstract}
Driven by their remarkable success in computer vision and inverse problem solving, score-based models are increasingly applied to wireless communications, where they show promise across a range of physical-layer tasks. However, despite this growing interest, the current literature often lacks a rigorous analysis of when score-matching offer a tangible advantage over traditional discriminative learning. This paper aims to address this gap through the use-case of channel estimation, a fundamental inverse problem in wireless systems. We present a theoretically grounded interpretation of score-based channel estimation through the lens of the perception–distortion tradeoff, identifying the conditions where score matching excels as well as its key limitations. 
In particular, by modeling downstream wireless tasks (e.g., capacity maximization) as \emph{functionals} of the channel estimation process, we quantify the \emph{excess risk} incurred by standard distortion-minimization approaches. 
Extensive numerical results show that under high predictive uncertainty, the large excess risk gap can be offset by score-based estimation, enabling near Bayesian-optimal precoding via the learned posterior, whereas in the low predictive uncertainty regime, discriminative distortion-minimization approaches are preferable due to lower complexity and more efficient use of model capacity.
\end{abstract}

\section{Introduction}

\acp{DM} \cite{pmlr-v37-sohl-dickstein15, ho2020denoising} --- or more generally \acp{SGM} \cite{song2020score} --- recently emerged as a dominant class of generative models, achieving state-of-the-art performance across a wide range of computer vision tasks and, more broadly, in solving ill-posed inverse problems. Their success stems from a combination of expressive modeling capacity, remarkable robustness to noise and model mismatch, and the ability to incorporate complex priors through learned data distributions. These properties have naturally motivated their rapid adoption in wireless communications, where inverse problems are pervasive and channel estimation remains a central --- and arguably the most critical --- task.

In the context of channel estimation, \acp{SGM} offer several appealing advantages over classical and discriminative learning-based estimators. By learning an implicit prior over channel realizations, DMs can regularize severely underdetermined estimation problems, exhibit robustness to noise and model mismatch, and operate effectively in regimes characterized by limited pilot resources or low \ac{SNR}. 
These properties have led a growing body of work to position generative channel estimation as a superior alternative to classical and discriminative learning-based estimators \cite{arvinte2022mimo, dmbasedchannelestimation, letafati2025diffusionmodelswirelesscommunications, 10930691, 10812969, 10946972, 10446413,strasser2025enhancements, wang2025radiodiff}, often reporting substantial empirical gains in distortion-based metrics (e.g., \ac{MSE}).

\begin{figure*}[t]
    \centering
    \begin{subfigure}{0.7\textwidth}
        \centering
\includegraphics[height=5.5cm]{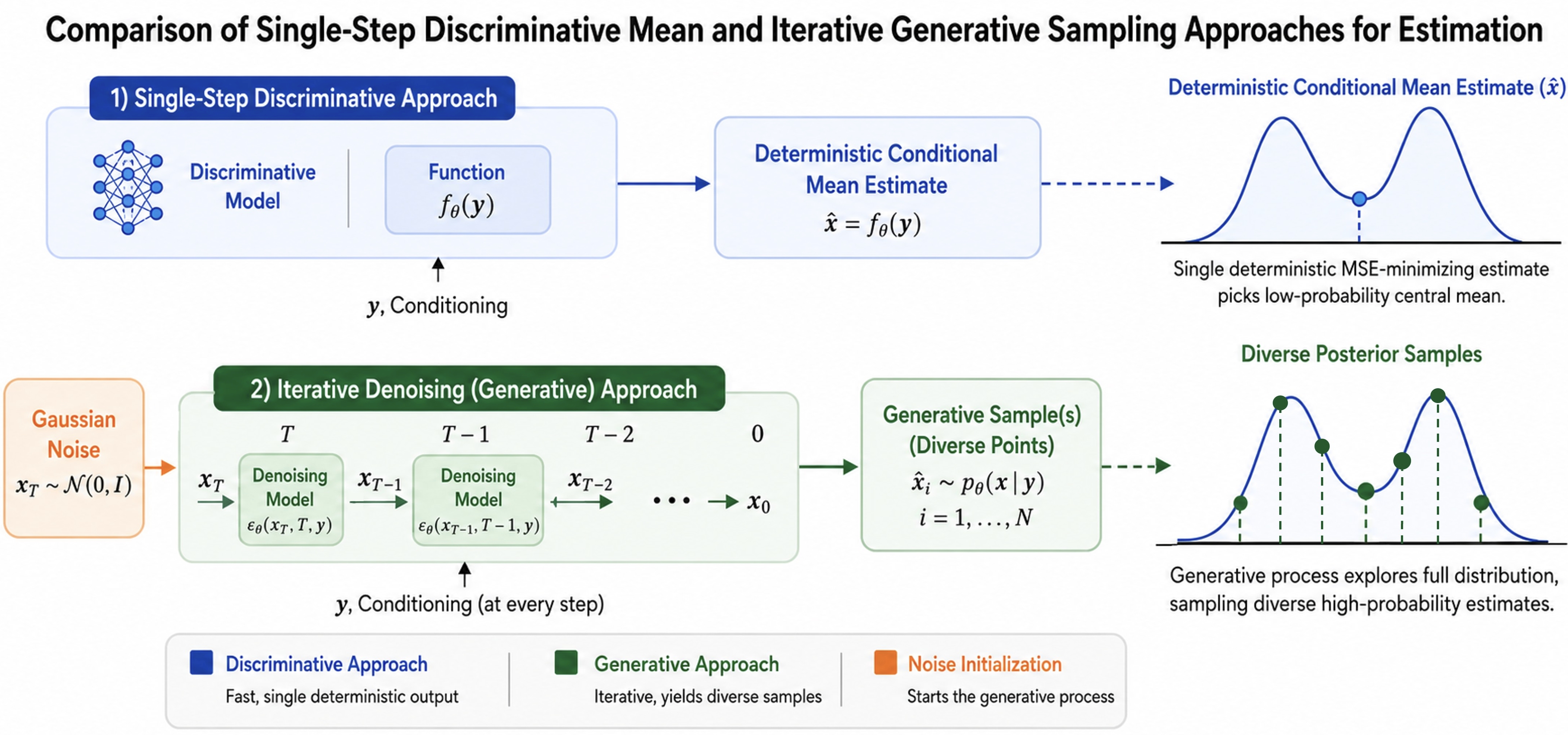}
        \caption{Comparison between Generative AI and Discriminative Learning frameworks.}
        \label{fig:overview_plot}
    \end{subfigure}\hspace{0.05\textwidth}
    \begin{subfigure}{0.2\textwidth}
        \centering
\includegraphics[height=5.5cm]{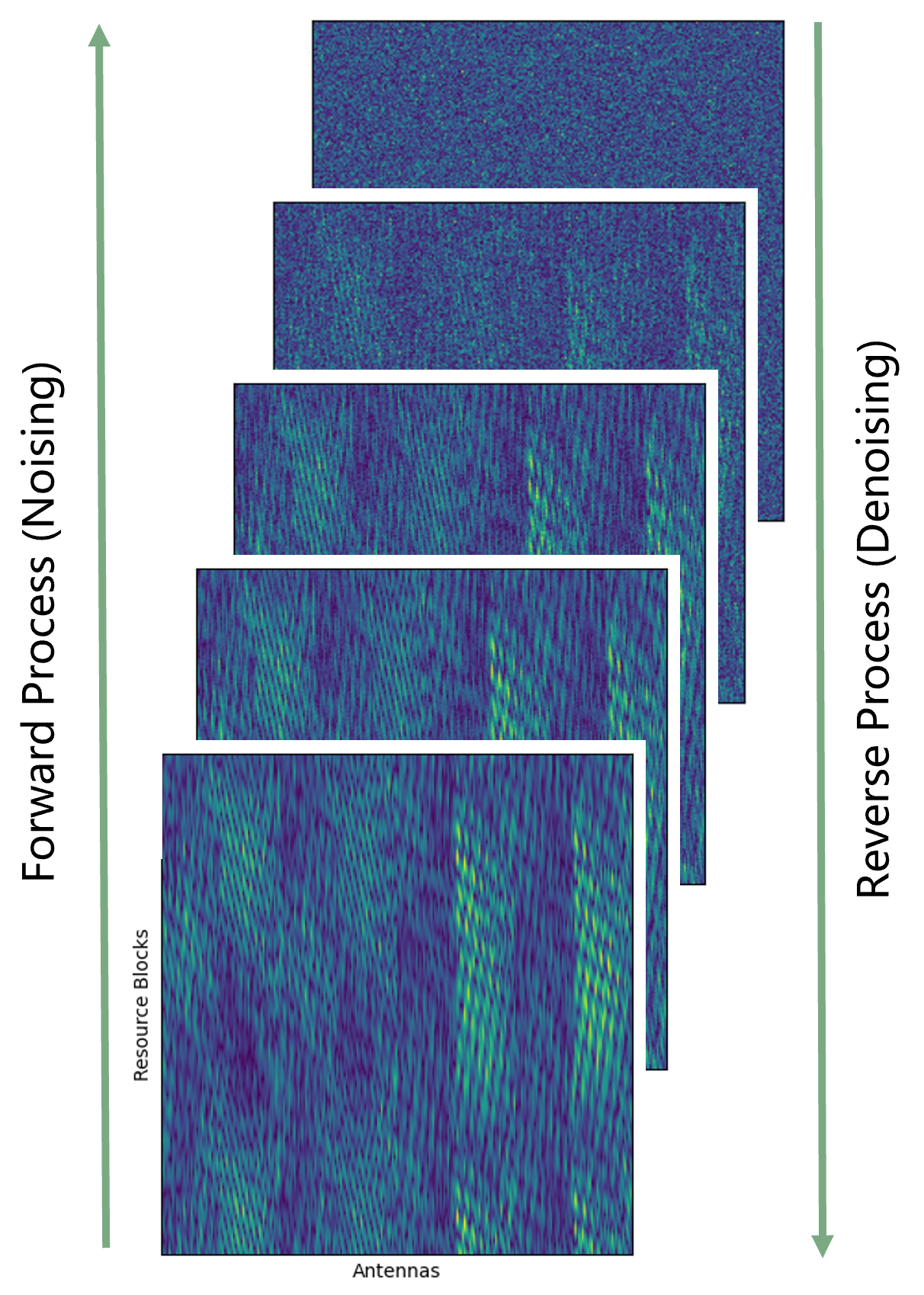}
        \caption{Diffusion process.}
        \label{fig:diffusion_process}
    \end{subfigure}
    \caption{\acp{SGM} provide samples from the posterior distribution through iterative refinement in the denoising process.}
    \label{fig:combined_overview_plots}
    \vspace{-0.35cm}
\end{figure*}

While the existing literature provides valuable insights, we argue that this prevailing focus on distortion minimization, overlooks the primary strength of generative score-based approaches: 
their capacity to produce highly plausible samples from learned conditional posterior distributions.
Unlike standard discriminative estimators that collapse to the \ac{CME}, generative models maintain full distributional awareness. As conceptually illustrated in Fig.~\ref{fig:combined_overview_plots}, the iterative denoising process prevents the model from outputting physically implausible, averaged hypotheses.
While the distinction between perceptual fidelity and distortion minimization is widely understood in computer vision through the perception-distortion tradeoff \cite{blau2018perception}, we aim to shed light on its significance within the wireless domain. This distinction's practical importance becomes clear when considering the effect of channel estimation on downstream communication tasks. Indeed, in wireless systems, the ultimate goal is seldom the accurate reconstruction of the channel itself. Instead, the focus is on optimizing channel-dependent functionals \cite{ha2025bayesian}, such as achievable rates, outage probabilities, or MIMO capacity. As expanded upon in the following sections, the discrepancy between distortion minimization and functional optimization is fundamentally explained by the Jensen gap \cite{CoverThomas2006}: except in the special case of linear task functionals, the two objectives do not coincide. 
\acp{SGM}, by providing access to well-matched posterior samples, are therefore suited to this problem, as they enable direct evaluation or optimization of task-relevant functionals in ways that point estimators inherently cannot.

In this work, we argue that the widespread characterization of \acp{SGM} as optimal direct estimators is both conceptually and practically limiting. Instead, we advocate for a reframing of their role as generative engines for \textit{channel-aware decision-making}, prioritizing downstream task utility over traditional distortion-based metrics.

\section{Related Work \& Contribution}

\subsection{Related Work}
\label{sec:related_work}

In computer vision, \acp{DM} \cite{pmlr-v37-sohl-dickstein15, ho2020denoising} and \acp{SGM} \cite{song2020score} have emerged as the state-of-the-art solution for generative tasks, replacing the long-standing dominance of \acp{GAN} in image synthesis \cite{dhariwal2021diffusion}. 
Their superiority stems from improved training stability and a lack of mode collapse, allowing them to effectively capture complex, high-dimensional data distributions \cite{nichol2021improved}.
Driven by this success, \acp{SGM} have been rapidly integrated into the wireless domain, proving effective across a range of applications as categorized in recent surveys \cite{10812969, fan2025generativediffusionmodelswireless, luong2025diffusionmodelsfuturenetworks, van2024generative}.
In the physical layer, these models have demonstrated their potential in channel modeling, estimation, and extrapolation \cite{10930691, 10946972, 11316498}, as well as signal detection \cite{10446413, 10888845}. 
\acp{DM} have also been shown to facilitate high-fidelity synthetic data generation for radio map estimation \cite{11066175, 10619829, 10843401} and digital twin construction \cite{wang2025radiodiff, 10906057}. 
Beyond estimation and synthesis, they support network management tasks --- for instance, by enabling trajectory sampling for reinforcement learning \cite{11391481} or acting as powerful network optimizers through inverse problem-solving \cite{liang2025diffusion}.
Furthermore, \acp{DM} have been explored as wireless foundation models \cite{fan2025generativediffusionmodelswireless}. This paradigm is particularly effective for \ac{MIMO} channel estimation, as pre-trained generative priors can be integrated with analytical likelihood functions for received pilot signals. Recent studies have demonstrated that such approaches achieve competitive reconstruction performance compared to both classical and other \ac{ML}-based methods \cite{arvinte2022mimo, 10930691, 10946972}.
Notably, since the generative prior is measurement-agnostic, these frameworks are inherently robust to sensing environment changes \cite{jalal2021robust, arvinte2022mimo}. This modularity allows the analytical likelihood to be exchanged at test-time to match specific hardware or pilot configurations without retraining the model.

Despite these promising results, a notable trend in the current literature is the reliance on distortion measures --- specifically \ac{MSE} --- as the primary evaluation metric for wireless \acp{DM} \cite{arvinte2022mimo, dmbasedchannelestimation, letafati2025diffusionmodelswirelesscommunications, 10930691, 10812969, 10946972, 10446413,strasser2025enhancements, wang2025radiodiff}. Studying distortion minimization properties for \acp{SGM} is motivated by their inherent advantages in generalization, robustness, and ``plug-and-play" versatility \cite{graikos2022diffusion}. 
However, we would argue that training via score-matching is inherently inefficient for distortion minimization, because the model must learn to denoise at every intermediate step of the forward process. 
As a consequence, significant model capacity is consumed by these intermediate steps, which do not necessarily improve the model's ability to perform the direct denoising task.
Instead, it is understood that \acp{SGM} can flexibly traverse the Pareto frontier of the perception-distortion tradeoff --- sacrificing distortion for perceptual quality \cite{wang2025traversing}. Notably, this has been also addressed in other wireless research, as in the denoinsing use case in \cite{fesl2024asymptotic}, which identifies the noise injected during the reverse process as a critical lever, allowing the \ac{DM} to prioritize either probabilistic accuracy or \ac{MSE} optimality. While variance scaling is one approach, several studies have shown that the number of inference steps acts as a primary lever for this balance. As shown in \cite{fabian2024diracdiffusiondenoisingincrementalreconstruction}, perceptual quality --- and thus distributional fidelity --- typically improves with additional steps, often at the cost of increased distortion. This inverse relationship likely explains why improved \ac{MSE} in wireless tasks is frequently reported for models with reduced steps \cite{wang2025radiodiff} or single-step predictors \cite{strasser2025enhancements}. 
However, in this edge case, a single-step inference score-based model becomes functionally equivalent to a discriminative one-step denoiser \cite{strasser2025enhancements}. This equivalence necessitates a direct comparison between the \ac{MSE} performance of diffusion-based models and identical backbones trained in a discriminative setting --- a baseline that remains underexplored in prior work.

Moving away from distortion metrics toward full distributional awareness more effectively highlights the unique strengths of generative AI in the wireless domain --- specifically for uncertainty-aware optimization where point-estimates prove insufficient. For instance, recent work demonstrates how diffusion models can be integrated into stochastic optimization frameworks to account for channel uncertainties \cite{10937314}. Similarly, in beamforming and precoding, utilizing the full distribution via \acp{VAE} enables robust strategies that outperform traditional methods under imperfect CSI \cite{11413254}. Even classical approaches like \acp{GMM} have shown that maintaining a probabilistic view of the channel is essential in challenging regimes, such as channel estimation with sparse pilots \cite{11202852}.

\subsection{Contribution}
Our contributions can be summarized as follows:

$\emph{i)}$ We provide a theoretically grounded interpretation of score-based channel estimation through the lens of the perception–distortion tradeoff. In doing so, we address a critical gap in the literature by identifying both the regimes where score matching excels and its fundamental limitations. Furthermore, we characterize the performance gap between generative score-matching and discriminative models when optimizing downstream task functionals, demonstrating that the resulting \textit{excess risk} arises naturally from Jensen’s inequality.
    
$\emph{ii)}$ Through extensive numerical experiments using MIMO precoding as a representative downstream task, we show how score-based estimation can outperform discriminative models in terms of achievable \ac{SE}, enabling approximate Bayesian-optimal precoding via the learned posterior distribution. Furthermore, we show how such approach can almost entirely match the performance of ad-hoc trained \ac{E2E} models on the MIMO capacity loss as an empirical upper bound not affected by excess risk.
    
$\emph{iii)}$ Finally, we draw a connection between the excess risk incurred by discriminative models and \ac{MI}. Using \ac{CFG}'s guidance scale parameter $\lambda$ as a \ac{MI} proxy, we show that the achievable spectral efficiency gap with score-matching is amplified in low-information regimes, where \acp{CME} suffer structural collapse. On the other hand, we show that in the high mutual-information regime, discriminative distortion-minimization approaches is typically preferable due to lower complexity and more efficient use of model capacity.

The remainder of this paper is organized as follows: Sections \ref{sec:preliminaries} and \ref{sec:perception_distortion_tradeoff} provide the theoretical foundations of score-based generative models and the perception-distortion tradeoff. Section \ref{sec:problem_formulation} details the application of these concepts to \ac{MIMO} channel estimation, before our proposed algorithmic solutions and experimental results are presented in Section \ref{sec:experiments}. Final conclusions are drawn in Section \ref{sec:conclusion}.

\section{Preliminaries}
\label{sec:preliminaries}
With the aim of making the paper self-contained, we review the theoretical background of \acp{DM}, focusing specifically on \acp{DDPM} and how they can be formally interpreted through the lens of \acp{SGM}.

\subsection{Denoising Diffusion Probabilistic Models}

\acp{DDPM} \cite{ho2020denoising} are a class of generative latent variable models that learn a data distribution $q(\mathbf{x}_0)$ by reversing a gradual noise-injection process. The framework is defined by two primary Markov chain processes: the forward diffusion process and the reverse generative process. \\

\subsubsection{Forward Diffusion Process}
The forward process, denoted by $q$, progressively transforms a data sample $\mathbf{x}_0 \sim q(\mathbf{x}_0)$ into latent variables $\mathbf{x}_1, \dots, \mathbf{x}_T$ by adding Gaussian noise according to a predefined variance schedule $\beta_1, \dots, \beta_T$:
\begin{equation}
    q(\mathbf{x}_t | \mathbf{x}_{t-1}) = \mathcal{N}(\mathbf{x}_t; \sqrt{1 - \beta_t} \mathbf{x}_{t-1}, \beta_t \mathbf{I}).
\end{equation}
A key property of this formulation is that it allows for sampling $\mathbf{x}_t$ at any arbitrary timestep $t$ directly from the original data $\mathbf{x}_0$. Defining $\alpha_t = 1 - \beta_t$ and $\bar{\alpha}_t = \prod_{s=1}^t \alpha_s$, the distribution of $\mathbf{x}_t$ conditioned on $\mathbf{x}_0$ is given by:
\begin{equation}
    q(\mathbf{x}_t | \mathbf{x}_0) = \mathcal{N}(\mathbf{x}_t; \sqrt{\bar{\alpha}_t} \mathbf{x}_0, (1 - \bar{\alpha}_t) \mathbf{I}).
\end{equation}
Using the reparameterization trick, this can be expressed as:
\begin{equation}\label{eq:forward_diffusion}
    \mathbf{x}_t = \sqrt{\bar{\alpha}_t} \mathbf{x}_0 + \sqrt{1 - \bar{\alpha}_t} \boldsymbol{\epsilon}, \quad \boldsymbol{\epsilon} \sim \mathcal{N}(0, \mathbf{I}).
\end{equation}
As $T \to \infty$, the final latent variable $\mathbf{x}_T$ converges to an isotropic Gaussian distribution.

\subsubsection{Reverse Generative Process}
The generative model aims to learn the reverse transition $p_\theta(\mathbf{x}_{t-1} | \mathbf{x}_t)$ to recover the data from noise. This reverse process is also modeled as a Gaussian transition:
\begin{equation}
    \label{eq:transition_kernel}
    p_\theta(\mathbf{x}_{t-1} | \mathbf{x}_t) = \mathcal{N}(\mathbf{x}_{t-1}; \boldsymbol{\mu}_\theta(\mathbf{x}_t, t), \boldsymbol{\Sigma}_\theta(\mathbf{x}_t, t)).
\end{equation}
The neural network is trained to estimate the mean $\boldsymbol{\mu}_\theta$, while the covariance is typically kept as a fixed constant $\boldsymbol{\Sigma}_\theta(\mathbf{x}_t, t) = \tilde{\beta}_t \mathbf{I}$ with $\tilde{\beta}_t = \frac{1-\bar{\alpha}_{t-1}}{1-\bar{\alpha}_t}\beta_t$. Rather than predicting the mean directly, the model is parameterized to predict the noise $\boldsymbol{\epsilon}$ added at timestep $t$, such that:
\begin{equation}
    \boldsymbol{\mu}_\theta(\mathbf{x}_t, t) = \frac{1}{\sqrt{\alpha_t}} \left( \mathbf{x}_t - \frac{\beta_t}{\sqrt{1 - \bar{\alpha}_t}} \boldsymbol{\epsilon}_\theta(\mathbf{x}_t, t) \right).
\end{equation}

\subsubsection{Training Objective and Parameterization}
The parameters $\theta$ are optimized by minimizing a simplified version of the variational upper bound on the negative log-likelihood, typically defined as the mean squared error in the noise space:
\begin{equation}
    \label{eq:mean_prediction}
    L_{\text{DM}}(\theta) = \mathbb{E}_{t, \mathbf{x}_0, \boldsymbol{\epsilon}} \left[ \| \boldsymbol{\epsilon} - \boldsymbol{\epsilon}_\theta(\mathbf{x}_t, t) \|^2 \right].
\end{equation}
where $t$ is sampled uniformly from $\{1, \dots, T\}$. While the standard DDPM objective focuses on predicting the noise $\boldsymbol{\epsilon}$, it is mathematically equivalent to predicting the original clean data $\mathbf{x}_0$. Given the forward diffusion relationship in (\ref{eq:forward_diffusion}), the clean signal can be estimated from the predicted noise as:
\begin{equation}
    \label{eq:x0_from_predicted_noise}
    \hat{\mathbf{x}}_0 = \frac{1}{\sqrt{\bar{\alpha}_t}} \left( \mathbf{x}_t - \sqrt{1 - \bar{\alpha}_t} \boldsymbol{\epsilon}_\theta(\mathbf{x}_t, t) \right).
\end{equation}
Conversely, the network can be designed to predict the clean signal $\mathbf{x}_0$ directly, from which the corresponding noise can be algebraically derived. As expanded upon in the next section, this objective allows the model to learn the score of the data distribution, enabling the generation of high-fidelity samples from pure Gaussian noise during inference.

\subsection{Score-based Generative Models}

\acp{SGM} \cite{song2019generative} take a complementary perspective to diffusion-based approaches: rather than explicitly modeling reverse-time conditionals, they learn the \emph{score function} of the noise-corrupted data distribution at each time step,
\begin{equation}
    s_\theta(\mathbf{x}_t, t) \approx \nabla_{\mathbf{x}_t} \log q(\mathbf{x}_t),
\end{equation}
i.e., the gradient of the log-density of \(\mathbf{x}_t\). Given an estimate of the score, samples can be generated by integrating a reverse-time \ac{SDE} or by performing Langevin dynamics.
Song et al.~\cite{song2020score} subsequently showed that \acp{DDPM} can be interpreted as a discrete-time approximation of such reverse-time \ac{SDE}s, thereby establishing diffusion models as discretized score-based generative models. Under Gaussian noise schedules, the DDPM noise-prediction objective is exactly equivalent to a weighted denoising score-matching loss, formalizing the connection between denoising and score estimation originally identified by Vincent~\cite{vincent2011connection}. In the DDPM parameterization \cite{ho2020denoising}, this relationship takes the explicit form
\begin{equation}
\label{eq:score-noise-equivalence}
\nabla_{\mathbf{x}_t} \log p(\mathbf{x}_t)
=
-\frac{\epsilon_\theta(\mathbf{x}_t, t)}{\sqrt{1 - \bar{\alpha}_t}}.
\end{equation}

\subsection{Conditional Diffusion Sampling}
Conditional generation aims to sample from the posterior $p(x_t | y)$, where $y$ represents a conditioning signal such as a class label, text prompt or in the context of wireless channel estimation, pilot observations. 
According to Bayes' rule, the gradient of the log-posterior (i.e., the score) decomposes into:
\begin{equation}
\nabla_{x_t} \log p(x_t | y) = \nabla_{x_t} \log p(x_t) + \nabla_{x_t} \log p(y | x_t).
\end{equation}

In classifier guidance, an external noisy classifier $p(y|x_t)$ provides the guidance gradient. 
Under an explicit measurement model, it is also possible to derive a guidance term in closed form from direct observations, as in \cite{arvinte2022mimo}. Substituting $\nabla_{\mathbf{x}_t} \log p(\mathbf{x}_t)$ as in Eq. \eqref{eq:score-noise-equivalence}, the noise estimate $\hat{\epsilon}$ with guidance scale $w$ becomes:
\begin{equation}
\hat{\epsilon} = \epsilon_\theta(x_t, t) - w \sqrt{1 - \bar{\alpha}_t} \nabla_{x_t} \log p(y | x_t).
\end{equation}
The negative sign ensures the update moves toward regions of higher classifier confidence. Alternatively, \ac{CFG} eliminates the need for an external model by jointly training on conditional and unconditional objectives. During training, $y$ is periodically replaced by a null token $\emptyset$ (e.g., a zero vector) to represent the unconditional state. At inference, the model extrapolates from the unconditional baseline:
\begin{equation}
    \label{eq:cfg}
    \hat{\epsilon} = \epsilon_\theta(x_t, t, \emptyset) + \lambda \left( \epsilon_\theta(x_t, t, y) - \epsilon_\theta(x_t, t, \emptyset) \right).
\end{equation}
This implicitly approximates the classifier gradient and serves as the input for the standard DDPM reverse step. \\

\section{MSE Optimality and the Perception-Distortion Tradeoff}
\label{sec:perception_distortion_tradeoff}

In this section, we study the statistical properties of \acp{SGM} through a Bayesian lens and contrast them with discriminative learning under the perception–distortion tradeoff. We show that while the \ac{CME} is optimal for MSE, it is generally suboptimal for nonlinear downstream functionals, with the resulting excess risk arising from Jensen’s inequality.

\begin{figure*}[htbp]
    \centering
    \includegraphics[width= 0.85 \linewidth]{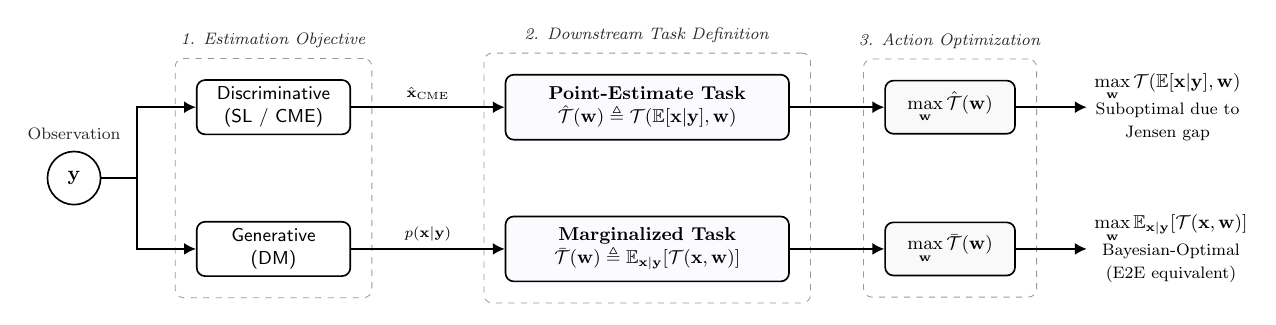}
    \caption{Distortion-optimal estimation leads to \textit{excess risk} for downstream tasks, resolvable through posterior marginalization.}
    \label{fig:flowchart_end_to_end}
    \vspace{-0.35cm}
\end{figure*}

\subsection{Conditional Mean Estimation vs. Posterior Sampling}

It is a well-known result from estimation theory that, under an \ac{MSE} distortion measure, the optimal estimator --- also referred to as the \ac{MMSE} estimator --- is given by the \ac{CME} \cite{kay1993fundamentals}:
\begin{equation}
    \hat{\mathbf{x}}_{\text{CME}} = \mathbb{E}[\mathbf{x} \mid \mathbf{y}] = \int_{\mathcal{X}} \mathbf{x} p(\mathbf{x} \mid \mathbf{y}) \, d\mathbf{x}.
\end{equation}

In a discriminative setting, a \ac{SL} model \( f_\theta \) is trained via \ac{ERM} under an \ac{MSE} loss to obtain a finite-sample approximation of the \ac{CME}:
\begin{equation}
    \min_\theta 
    \mathbb{E}_{(\mathbf{x},\mathbf{y}) \sim \mathcal{D}(\mathbf{x},\mathbf{y})}
    \big[ \| \mathbf{x} - f_\theta(\mathbf{y}) \|^2 \big],
\end{equation}
where \( \mathcal{D}(\mathbf{x},\mathbf{y}) \) denotes the empirical training distribution. 
While \ac{ERM} aims for the \ac{CME}, factors such as finite data and limited model capacity mean that $f_\theta$ only approximates the optimal estimator, reaching it only in the asymptotic limit under consistent \ac{ERM} with sufficiently expressive models, as characterized by the universal approximation theorem \cite{hornik1989multilayer}.

In contrast, an \ac{SGM} learns an explicit representation of the data prior $p(\mathbf{x})$, the conditional distribution $p(\mathbf{x}\mid\mathbf{y})$, or both, as in \ac{CFG}. Upon observation of a conditioning signal $\mathbf{y}$, inference consists of drawing a stochastic sample $\mathbf{x}' \sim p_\theta(\mathbf{x}\mid\mathbf{y})$. Because $\hat{\mathbf{x}}_{\text{CME}}$ is the unique minimizer of the MSE, any individual sample $\mathbf{x}'$ from the \ac{SGM} will necessarily yield a higher MSE than the CME. The total error of a sample can be decomposed into the MMSE and the sampling variance:
\begin{equation}
    \begin{aligned}
        \mathbb{E}[\|\mathbf{x} - \mathbf{x}'\|^2] &= \underbrace{\mathbb{E}[\|\mathbf{x} - \hat{\mathbf{x}}_{\text{CME}}\|^2]}_{\text{MMSE}} + \underbrace{\mathbb{E}[\|\mathbf{x}' - \hat{\mathbf{x}}_{\text{CME}}\|^2]}_{\text{Sampling Variance}}
    \end{aligned}
\end{equation}
The second term represents the inherent tradeoff between distributional fidelity and point-wise accuracy. While this sampling variance facilitates the generation of plausible channel realizations, it prevents a single sample from matching the \ac{MSE} of the \ac{CME}.

Although a single posterior sample is not MSE-optimal, an \ac{SGM} can, in principle, recover the approximate CME. For instance, through Monte Carlo averaging over samples from the learnt posterior:
\begin{equation}
\label{eq:MCA_diff}
    \hat{\mathbf{x}}_{\text{CME}} \approx \frac{1}{N} \sum_{i=1}^N \mathbf{x}^{(i)}, \quad \mathbf{x}^{(i)} \sim p_\theta(\mathbf{x}|\mathbf{y})
\end{equation}
As $N \to \infty$, this average converges to $\mathbb{E}_{p_\theta(\mathbf{x}\mid\mathbf{y})}[\mathbf{x}]$. Therefore, generative models inherently generalize standard regression, as the minimum-risk point estimate can be derived directly from the full posterior. Beyond Monte Carlo averaging, efficient sampling strategies can force an \ac{SGM} to yield \ac{MSE}-optimal estimates for pure denoising tasks. Whether through modified reverse trajectories \cite{fesl2024asymptotic} or single-step inference \cite{strasser2025enhancements}, these approaches share a common mechanism: they systematically remove stochasticity from the generation process. By effectively collapsing the diverse posterior into a single conditional mean, these methods achieve minimal distortion precisely by sacrificing the model's generative properties. This dynamic is a direct manifestation of the \textit{perception–distortion tradeoff} \cite{blau2018perception}, which formalizes the fundamental incompatibility between distortion-optimal estimation (e.g., \ac{MMSE}) and perceptually faithful reconstruction (e.g., maximizing log-likelihood).

\subsection{Functional Suboptimality \& Excess Risk}
\label{sec:functionals_end2end}
Let us define $\mathbf{x} \in \mathcal{X}$ as a target variable (e.g., the channel state), $\mathbf{y} \in \mathcal{Y}$ as the observable data (e.g., received pilots), and $\hat{\mathbf{x}} = f(\mathbf{y})$ as the estimation of $\mathbf{x}$ given $\mathbf{y}$. Let us further define $\mathbf{w} = g(\hat{\mathbf{x}})$ as a task-specific action (e.g., precoding matrix selection) contingent upon the estimate. In many wireless systems, $\hat{\mathbf{x}}$ is treated as an intermediate latent representation used to solve for the final action $\mathbf{w}$, supporting diverse downstream tasks. The system performance is expressed as a functional $\mathcal{T}(\mathbf{x}, \mathbf{w})$, representing the utility achieved when action $\mathbf{w}$ is applied to the true state $\mathbf{x}$. We define the Bayes risk as the expected loss over the joint distribution:
\begin{equation}
R(f, g) = \mathbb{E}_{p(\mathbf{x}, \mathbf{y})} \big[ -\mathcal{T}(\mathbf{x}, g(f(\mathbf{y}))) \big]~.
\end{equation}
The task-optimal action $\mathbf{w}^{\star}$ for a given observation $\mathbf{y}$ is found by maximizing the posterior expected utility:
\begin{equation}
\mathbf{w}^{\star}(\mathbf{y}) = \arg\max_{\mathbf{w}} \mathbb{E}_{p(\mathbf{x} \mid \mathbf{y})} \big[ \mathcal{T}(\mathbf{x}, \mathbf{w}) \big]    
\end{equation}
In practice, the standard discriminative approach relies on the \ac{CME} $\hat{\mathbf{x}} = \mathbb{E}[\mathbf{x} \mid \mathbf{y}]$, and selects an action $\mathbf{w}_{\text{cme}} = g(\hat{\mathbf{x}})$ that is optimal only for that specific mean state. However, unless $\mathcal{T}$ is affine with respect to $\mathbf{x}$, this results in a suboptimal approach. Because the true performance is governed by the ground-truth $\mathbf{x}$, Jensen's inequality \cite{CoverThomas2006} reveals a systemic gap between the utility of the average state and the average utility of the state:
\begin{equation}
\mathcal{T}(\mathbb{E}[\mathbf{x} \mid \mathbf{y}], \mathbf{w}) \neq \mathbb{E}_{\mathbf{x} \mid \mathbf{y}}[\mathcal{T}(\mathbf{x}, \mathbf{w})]~.   
\end{equation}
This discrepancy results in an \textit{excess risk}. Relying on the CME provides an action $\mathbf{w}_{\text{cme}}$ that is optimized for a representative point rather than the actual distribution, leading to an expected performance that is offset by $\Delta \geq 0$ from the true optimal solution $\mathbf{w}^{\star}$. While this gap can be closed by \ac{E2E} frameworks that learn a direct mapping $\mathbf{w} = g_\theta(\mathbf{y})$, such approaches often sacrifice modularity and generality.
The Bayesian-optimal decision, which avoids this excess risk, is achieved by marginalizing out the unknown channel over its posterior distribution $p(\mathbf{x} \mid \mathbf{y})$:
\begin{equation}
\mathbf{w}^* = \arg\max_{\mathbf{w}} \int \mathcal{T}(\mathbf{x}, \mathbf{w}) p(\mathbf{x} \mid \mathbf{y}) d\mathbf{x}~.
\end{equation}
\acp{SGM} provide a powerful mechanism to overcome the intractability of this integral. By enabling efficient sampling from the posterior, \acp{SGM} allow us to approximate the objective via Monte Carlo integration:
\begin{equation}
    \mathbf{w}^* \approx \arg\max_{\mathbf{w}} \frac{1}{N} \sum_{i=1}^N \mathcal{T}(\mathbf{x}^{(i)}, \mathbf{w}), \quad \mathbf{x}^{(i)} \sim p_\theta(\mathbf{x} \mid \mathbf{y}).
\end{equation}
As demonstrated in Appendix A, maximizing the posterior expected utility via a generative model is mathematically equivalent to the optimal solution of a fully E2E framework. Consequently, provided the downstream optimization over $\mathbf{w}$ is solvable, the generative approach theoretically closes the performance gap to E2E systems, as illustrated in Fig. \ref{fig:flowchart_end_to_end}. This yields a profound architectural advantage: \acp{SGM} function as \textit{general-purpose} channel estimators that preserve the modularity of classical communication pipelines while simultaneously unlocking the potential of task-specific, \ac{E2E} learning. This theoretical equivalence is further validated numerically in Section \ref{sec:experiments}, where we demonstrate that the proposed generative approach successfully closes the performance gap between classical modular architectures and fully \ac{E2E}-trained systems.

\section{Problem Formulation}
\label{sec:problem_formulation}
We extend the principles established in Section \ref{sec:perception_distortion_tradeoff} to the task of \ac{MIMO} precoding. Our analysis focuses specifically on a 5G \ac{NR} \ac{TDD} system, examining \ac{DL} precoding based on \ac{UL} \ac{SRS} measurements --- a use-case of particular interest within 3GPP R20/21, e.g., \cite{huawei2026csi}.

\subsection{MIMO Channel Estimation}

We consider a point-to-point \ac{MIMO} system with $N_{\text{tx}}$ transmit and $N_{\text{rx}}$ receive antennas, transmitting over $N_f$ frequency subcarriers. 
To capture the temporal evolution of the channel, we observe the system over a sequence of $N_t$ discrete past time steps, $\{t_1, t_2, \dots, t_{N_t}\}$. 
Under practical 5G \ac{NR} configurations, \ac{SRS} transmissions employ an interleaved comb structure in the frequency domain and hop across different sub-bands.

Consequently, at each time step $t_m$, $N_p$ pilot symbols are transmitted only on a specific subset $P_m \subset \{1, \dots, N_f\}$ of subcarriers. 
The \ac{UL} received pilot signal tensor at the \acs{BS} at time $t_m$, denoted as $\mathbf{Y}_m^{(p)} \in \mathbb{C}^{N_{\text{rx}} \times N_p \times |P_m|}$, is modeled as:
\begin{equation}
    \mathbf{Y}_m^{(p)} = \mathcal{A}_{\mathbf{P}}^{(m)}(\mathbf{H}_{m}) + \mathbf{N}_m, \quad m = 1, \dots, N_t,
\end{equation}
where $\mathbf{H}_{m} \in \mathbb{C}^{N{\text{rx}} \times N_{\text{tx}} \times N_f}$ is the unknown true channel tensor at the past time $t_m$ (with the \ac{UL} channel being its transpose due to \ac{TDD} reciprocity), and $\mathbf{P}_m \in \mathbb{C}^{N_{\text{tx}} \times N_p \times |P_m|}$ is the pilot matrix. The tensor $\mathbf{N}_m \in \mathbb{C}^{N_{\text{rx}} \times N_p \times |P_m|}$ contains additive 
white Gaussian noise with i.i.d.\ entries $n_{i,j,k} \sim \mathcal{CN}(0, \eta^2)$, and $\mathcal{A}_{\mathbf{P}}^{(m)}$ denotes the time-specific linear pilot observation mapping, 
defined such that $\mathcal{A}_{\mathbf{P}}^{(m)}(\mathbf{H}_{m})_{:,:,k} = \mathbf{H}_{m,:,:,k}\mathbf{P}_{m,:,:,k}$ for $k \in P_m$. 
Based on the history of sparse, sequentially hopped observations, the ultimate goal is to predict the full wideband channel to perform \ac{DL} precoding at the current transmission time $t \geq t_{N_t}$. 
To align our physical model with the generative framework in Section \ref{sec:perception_distortion_tradeoff}, we define the observation history $\mathbf{y}$ and the ground-truth target state $\mathbf{x}$ as:
\begin{equation}
    \begin{aligned}
        \mathbf{y} \leftrightarrow \mathbf{Y}^{(p)} &\triangleq {\mathbf{Y}_1^{(p)}, \mathbf{Y}_2^{(p)}, \dots, \mathbf{Y}_{N_t}^{(p)}}, \\
        \mathbf{x} \leftrightarrow \mathbf{H} &\triangleq \mathbf{H}_{t} \in \mathbb{C}^{N_{\text{rx}} \times N_{\text{tx}} \times N_f}.
    \end{aligned}
\end{equation}

\begin{table*}[t]
    \centering
\resizebox{0.995\textwidth}{!}{
        \begin{tabular}{@{} l l l @{\hspace{8ex}} l l l @{}}
            \toprule
            \textbf{Category} & \textbf{Parameter} & \textbf{Value} & \textbf{Category} & \textbf{Parameter} & \textbf{Value} \\
            \midrule
            
\multirow{4}{*}{\textbf{Scenario Setup}} 
            & Environment & 3GPP UMa & 
            \multirow{4}{*}{\textbf{Node Config.}} 
            & BS Height & $25.0$ m \\
            
& Carrier Frequency & $3.7$ GHz &
            & UT Height & $1.5$ m \\
            
& Inter-Site Distance & $300$ m &
            & UT Vel. (Low Mob.) & $3$ km/h (Out) / $0.5$ km/h (In) \\
            
& SNR Range & UL: [-10, 10], DL: UL + $7$dB &
            & UT Vel. (High Mob.) & $18$ km/h (Out) / $3$ km/h (In) \\
            
\cmidrule(r){1-3} \cmidrule(l){4-6} 
            
\multirow{4}{*}{\textbf{Antenna Arrays}} 
            & BS Array & $4 \times 8$, Dual-Pol. ($64$ elem.) &
            \multirow{4}{*}{\textbf{Grid \& Pilots}} 
            & Subcarrier Spacing & $30$ kHz \\
            
& UT Array & $1 \times 2$, Dual-Pol. ($4$ elem.) &
            & PRB Allocation & $272$ PRBs ($100$ MHz) \\
            
& BS Elem. Spacing & $2.01\lambda$ (V), $0.5\lambda$ (H) &
            & SRS Config. & $40$ ms Period., $17$ Hops ($16$ PRBs/hop) \\
            
& Antenna Pattern & 3GPP TR 38.901 &
            & Prediction Target & Full band, $20$ ms after last SRS \\
            
            \bottomrule
        \end{tabular}
    }
    \caption{Simulation Scenario and System Parameters}
    \label{tab:sim_parameters}
\end{table*}

\subsection{Precoding as the Downstream Task}
\label{sec:precoding_task}

To evaluate the downstream impact of the channel estimation strategy, we consider wideband \ac{DL} spatial multiplexing. At the current transmission time, the \ac{BS} uses the predicted channel to map data streams onto its $N_{\text{tx}}$ antennas. 
For a specific subcarrier $k$, let $\mathbf{H}_k \in \mathbb{C}^{N_{\text{rx}} \times N_{\text{tx}}}$ denote the $k$-th subcarrier slice of our true target channel tensor $\mathbf{H}$. 
The received \ac{DL} data signal at the \acs{UE} is modeled as:
\begin{equation}
    \mathbf{Y}^{(s)}_k = \mathbf{H}_k \mathbf{W}_k \mathbf{S}_k + \mathbf{N}_k, \quad k = 1, \dots, N_f,
\end{equation}
where $\mathbf{W}_k \in \mathbb{C}^{N_{\text{tx}} \times N_L}$ is the precoding matrix, $\mathbf{S}_k$ contains the transmitted data symbols, 
and $\mathbf{N}_k$ is the \ac{DL} additive noise.
The ultimate objective of the \ac{BS} is to design the wideband precoder $\mathbf{W} = \{\mathbf{W}_k\}_{k=1}^{N_f}$ to maximize the achievable \ac{MIMO} sum-capacity:
\begin{equation} 
    \label{eq:capacity_task}
    \mathcal{T}(\mathbf{H}, \mathbf{W}) \triangleq \sum_{k=1}^{N_f} \log_2\det\!\left( \mathbf{I}_{N_\text{rx}} + \frac{1}{\eta^2} \mathbf{H}_k \mathbf{W}_k \mathbf{W}_k^{H} \mathbf{H}_k^{H} \right),
\end{equation}
where power allocation is absorbed into $\mathbf{W}_k$ under a sum-power constraint. With perfect \ac{CSI}, the optimal precoder is $\mathbf{W}_k = \mathbf{V}_k$, obtained from the \ac{SVD} $\mathbf{H}_k = \mathbf{U}_k\mathbf{\Sigma}_k\mathbf{V}_k^H$. 
However, inferring $\mathbf{H}$ from sparse, outdated \ac{UL} observations $\mathbf{Y}^{(p)}$ introduces uncertainty, which can not be resolved through the \ac{CME} estimate, but requires proper marginalization of the capacity functional over the posterior $p(\mathbf{H} \mid \mathbf{Y}^{(p)})$:
\begin{equation}
    \label{eq:bayesian_precoding_integral}
    \mathbf{W}^* = \arg\max_{\mathbf{W}} \int \mathcal{T}(\mathbf{H}, \mathbf{W}) p(\mathbf{H} \mid \mathbf{Y}^{(p)}) \, d\mathbf{H}.
\end{equation}

\section{Algortihms}
\label{sec:algorithms}
Since \eqref{eq:bayesian_precoding_integral} is analytically intractable, we approximate the optimal precoding matrix using Monte Carlo integration based on the learned posterior from the \ac{DM}.
\begin{equation}
\label{eq:monte_carlo_precoding}
\mathbf{W}^* \approx \arg\max_{\mathbf{W}} \frac{1}{N} \sum_{i=1}^{N} \mathcal{T}(\mathbf{H}^{(i)}, \mathbf{W}), \quad \mathbf{H}^{(k)} \sim p_\theta(\mathbf{H} \mid \mathbf{Y}^{(p)})
\end{equation}
The optimization problem in \eqref{eq:monte_carlo_precoding} can be cast as a \emph{manifold-constrained} optimization, where each feasible $\mathbf{W} \in \mathbb{C}^{N_{\text{tx}}\times N_L}$ has orthonormal columns. Consequently, the optimal solution lies on the complex Stiefel manifold \cite{absil2008optimization}, and standard Riemannian optimization techniques, such as the conjugate gradient method can be employed to solve it.

However, for the \ac{DM}-based precoding approach, we consider the solution to the related MIMO objective of maximizing the received energy:
\begin{equation}
    \tilde{\mathcal{T}}(\mathbf{H}, \mathbf{W}) \triangleq \sum_{k=1}^{N_f} \|\mathbf{H}_k \mathbf{W}_k\|_F^2 = \sum_{k=1}^{N_f} \text{tr}\!\left(\mathbf{W}_k^H \mathbf{H}_k^H \mathbf{H}_k \mathbf{W}_k\right),
\end{equation}
This objective is preferred because: $\emph{(i)}$ it affords a closed-form solution for the Bayesian precoder, \emph{(ii)} it offers lower implementation complexity compared to the conjugate gradient method, and \emph{(iii)} it empirically produces precoding matrices that yield nearly identical achievable SE to the original capacity task \eqref{eq:capacity_task}. By the linearity of the expectation and trace operators, the optimal Bayesian precoder for a given subcarrier $k$ is then given by the $N_L$ dominant eigenvectors of the expected posterior Gramian defined as $\mathbf{R}_k = \mathbb{E}_{\mathbf{H} \mid \mathbf{Y}^{(p)}} [\mathbf{H}_k^H \mathbf{H}_k]$ --- see Appendix \ref{app:rayleigh_ritz} for details.
This yields Algorithm \ref{alg:generative_precoding}, where the \ac{DM} constructs the expected Gramian $\mathbf{R}_k$ through Monte Carlo approximation, drawing a set of samples from the posterior.

\begin{algorithm}[t]
\caption{Precoder Design via Generative Learning (DM)}
\label{alg:generative_precoding}
\begin{algorithmic}[1]
\REQUIRE Sparse observation history $\mathbf{Y}^{(p)}$, sample count $N$.
\STATE Sample the channel posterior: $\{\mathbf{H}^{(i)}\}_{i=1}^N \sim p_{\theta}(\mathbf{H} \mid \mathbf{Y}^{(p)})$.
\FOR{each subcarrier $k = 1, \dots, N_f$}
    \STATE Construct the expected Gramian: \\ $\hat{\mathbf{R}}_k = \frac{1}{N} \sum_{i=1}^N (\mathbf{H}_{k}^{(i)})^H \mathbf{H}_{k}^{(i)}$.
    \STATE Compute $\mathbf{W}_k^*$ as the $N_L$ dominant eigenvectors of $\hat{\mathbf{R}}_k$.
\ENDFOR
\RETURN Wideband precoder $\mathbf{W}^* = \{\mathbf{W}_k^*\}_{k=1}^{N_f}$.
\end{algorithmic}
\end{algorithm}

\section{Experiments}
\label{sec:experiments}

\begin{figure}\includegraphics[width=\linewidth]{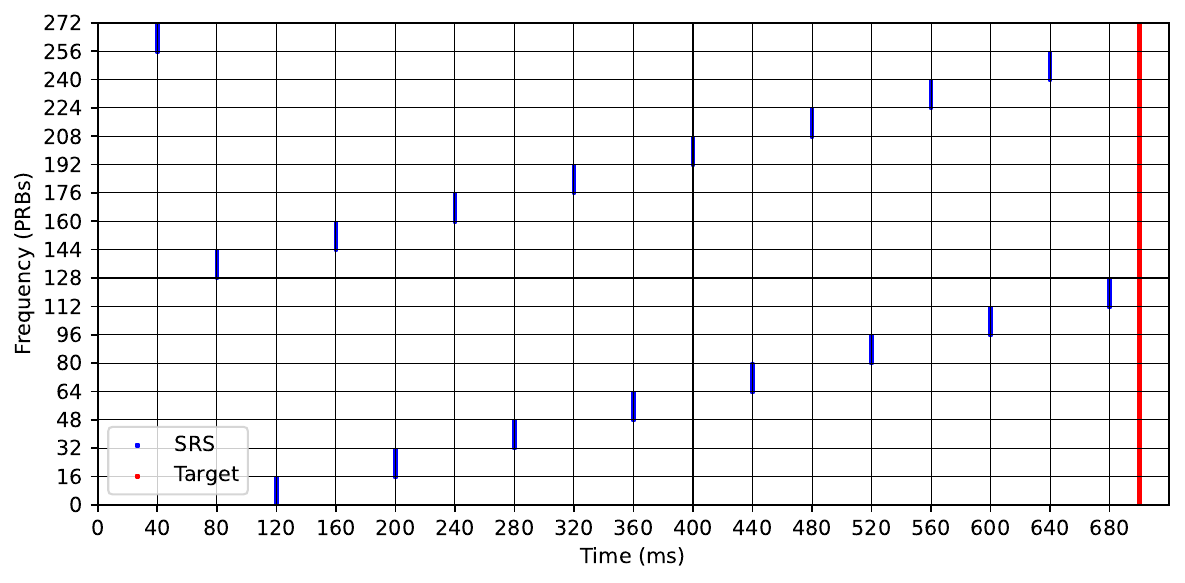}
    \caption{Time-frequency allocation of SRS pilots and target channel.}
    \label{fig:srs_pattern}
    \vspace{-0.35cm}
\end{figure}

\begin{figure}[t!]
    \centering
    \vspace{-0.65cm}
    \includegraphics[width= 0.95 \linewidth]{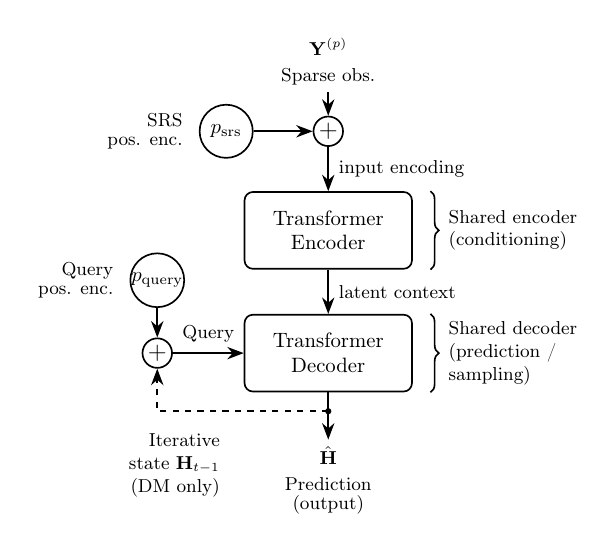}
    \vspace{-0.65cm}
    \caption{Used backbone model for DM and SL mode.}
    \label{fig:transformer_backbone}
    \vspace{-0.25cm}
\end{figure}

To comprehensively evaluate the utility of generative \acp{DM} against discriminative baselines, we conduct a series of experiments designed to assess \ac{MSE} performance, characterize the perception-distortion tradeoff, and quantify the incurred excess risk of discriminative solutions in MIMO precoding.

Our simulations adopt the system parameter configurations detailed in Table \ref{tab:sim_parameters} and are implemented using the Sionna open source library \cite{hoydis2022sionna}. Specifically, we consider two reference scenario of increasing difficulty: \textit{i)} \textit{low mobility} ($3$ / $0.5$~km/h) and \textit{ii)} \textit{high mobility} ($18$ / $3$~km/h). 
In both settings, the \ac{BS} conducts wideband \ac{UL} channel estimation based on a sequence of $N_t=17$ \ac{SRS} hops with a 40~ms periodicity, as illustrated in Fig. \ref{fig:srs_pattern}. The predictive challenge is defined by a fixed temporal gap of 20~ms between the final \ac{SRS} observation and the target wideband channel $\mathbf{H}$. 
Assuming \ac{UL}-\ac{DL} channel reciprocity, the predicted channel is directly utilized for the subsequent \ac{DL} precoding task. This setup represents a realistic channel aging scenario, currently considered in several 3GPP study items (e.g., \cite{huawei2026csi}) where the \ac{BS} is forced to perform downstream optimization based on severely outdated measurements.

For a controlled comparison, both the discriminative \ac{SL} model and the generative \ac{DM} share the identical Transformer backbone sketched in Fig. \ref{fig:transformer_backbone}, which employs an encoder to process the history of sparse observations into a latent context. In the discriminative \ac{SL} model --- trained via \ac{MSE} 
to approximate the \ac{CME} --- the decoder generates a single point-estimate using a query sequence consisting purely of target positional encodings. Conversely, during \ac{DM} inference, the encoder 
remains fixed while the decoder runs in an iterative loop where the input combines target positional encodings with the current intermediate denoised state. This iterative refinement allows the 
generative model to recover the structural details of the channel that are inherently lost in the averaging process of the discriminative \ac{SL} model trained via \ac{MSE}.
We refer the interested reader to Appendix \ref{app:model_description} for a detailed description of the model architecture and training procedure.

\subsection{Empirical Perception vs. Distortion Analysis}
\label{sec:experiments_perception}
{First, we conduct a set of experiments to characterize the implications of the perception-distortion tradeoff in the context of wireless channel estimation.

Table \ref{tab:results_mse} presents the \ac{NMSE} and \ac{GCS} metrics for the evaluated models\footnote{\ac{NMSE} is defined as $||\mathbf{H} - \hat{\mathbf{H}}||_F^2 / ||\mathbf{H}||_F^2$, while \ac{GCS} is computed as $\frac{|\langle \mathbf{H}, \hat{\mathbf{H}} \rangle|}{\|\mathbf{H}\|_F \|\hat{\mathbf{H}}\|_F}$, measuring the alignment independent of scaling.}. Notably, despite sharing the exact same neural network backbone, the discriminative \ac{SL} model consistently and substantially outperforms the posterior samples generated by the \ac{DM} across both mobility regimes\footnote{Table \ref{tab:results_mse} also includes the \ac{SRS} baseline, which is obtained by collecting the wideband estimate from the observations $\mathbf{Y}^{(p)}$ without any processing.}.
While this substantial performance gap might initially seem counterintuitive given the strong focus on \ac{DM} \ac{MSE} capabilities in the literature (as discussed in Section~\ref{sec:related_work}), it is a perfect physical manifestation of the perception-distortion tradeoff. By training via \ac{ERM}, the discriminative model is strictly encouraged to minimize distortion. Conversely, the \ac{DM} learns the true underlying data distribution, inherently prioritizing perceptual quality (i.e., physical channel realism) over absolute point-wise accuracy during inference.

\begin{table}\centering
    \resizebox{0.95\linewidth}{!}{
    \begin{tabular}{@{} l l c c @{}}
        \toprule
        \textbf{Scenario} & \textbf{Model} & \textbf{NMSE} $\downarrow$ & \textbf{GCS} $\uparrow$ \\
        \midrule
        \multirow{3}{*}{Low Mobility} 
        & SRS Baseline & $1.866$ & $0.381$ \\
        & SL Model & $\mathbf{0.431}$ & $\mathbf{0.750}$ \\
        & DM Model & $0.984$ & $0.602$ \\
        \midrule
        \multirow{3}{*}{High Mobility} 
        & SRS Baseline & $1.994$ & $0.276$ \\
        & SL Model & $\mathbf{0.809}$ & $\mathbf{0.534}$ \\
        & DM Model & $1.581$ & $0.405$ \\
        \bottomrule
    \end{tabular}
    }
    \caption{Channel Estimation Performance (Distortion).}
    \label{tab:results_mse}
    \vspace{-0.25cm}
\end{table}

\begin{figure*}[t] \centering
    \begin{subfigure}[t]{0.32\linewidth}
        \centering
        \includegraphics[width=\linewidth]{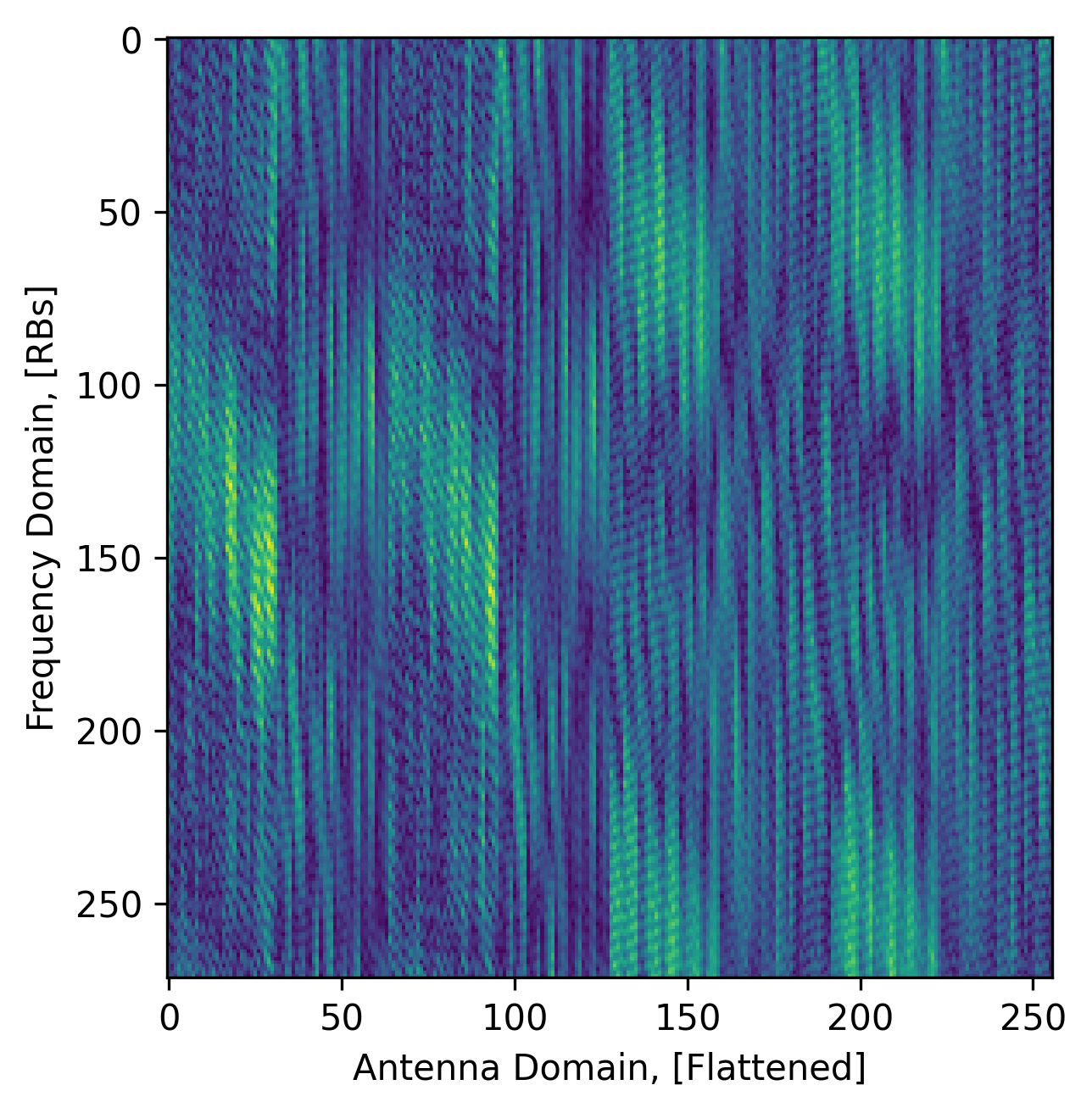}
        \caption{Ground Truth - \textit{Low Mobility}}
        \label{fig:perception-distortion-a}
    \end{subfigure}\hfill
    \begin{subfigure}[t]{0.32\linewidth}
        \centering
        \includegraphics[width=\linewidth]{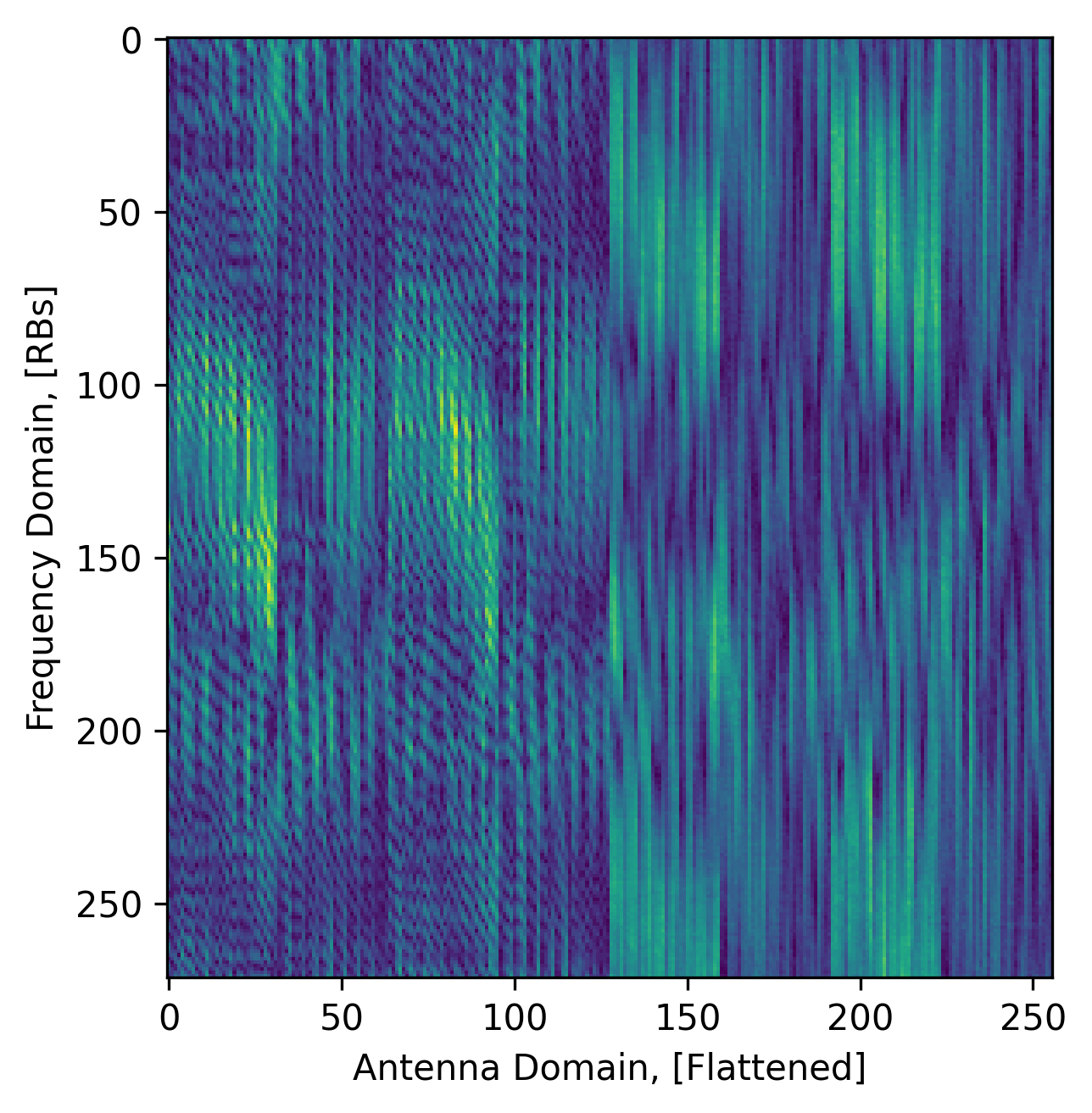}
        \caption{DM Estimate, NMSE=0.41, CS=0.80.}
        \label{fig:perception-distortion-b}
    \end{subfigure} \hfill
    \begin{subfigure}[t]{0.32\linewidth}
        \centering
        \includegraphics[width=\linewidth]{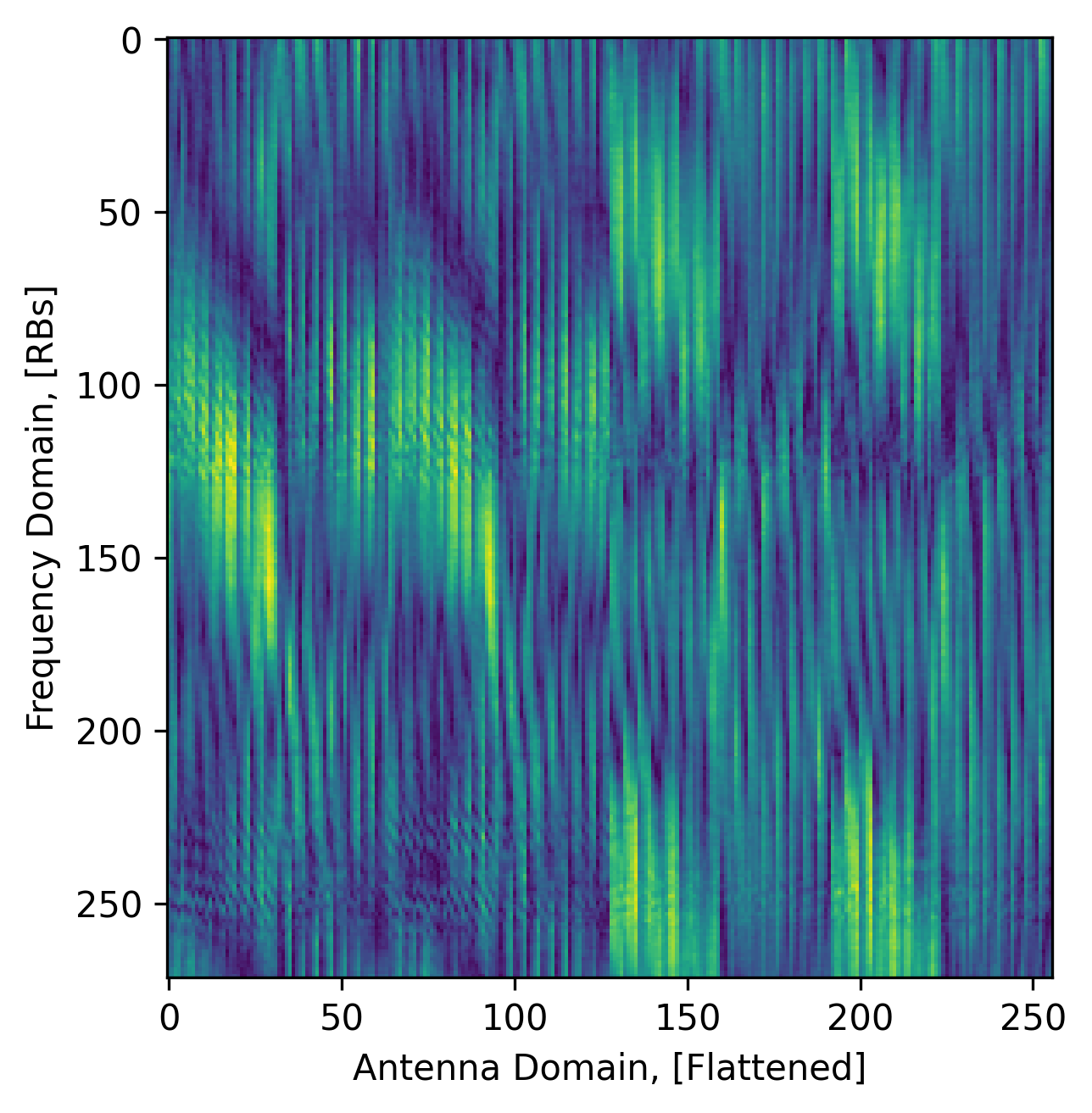}
        \caption{SL Estimate, NMSE=0.20, CS=0.90.}
        \label{fig:perception-distortion-c}
    \end{subfigure} \vspace{0.25cm} \\
    \begin{subfigure}[t]{0.32\linewidth}
        \centering
        \includegraphics[width=\linewidth]{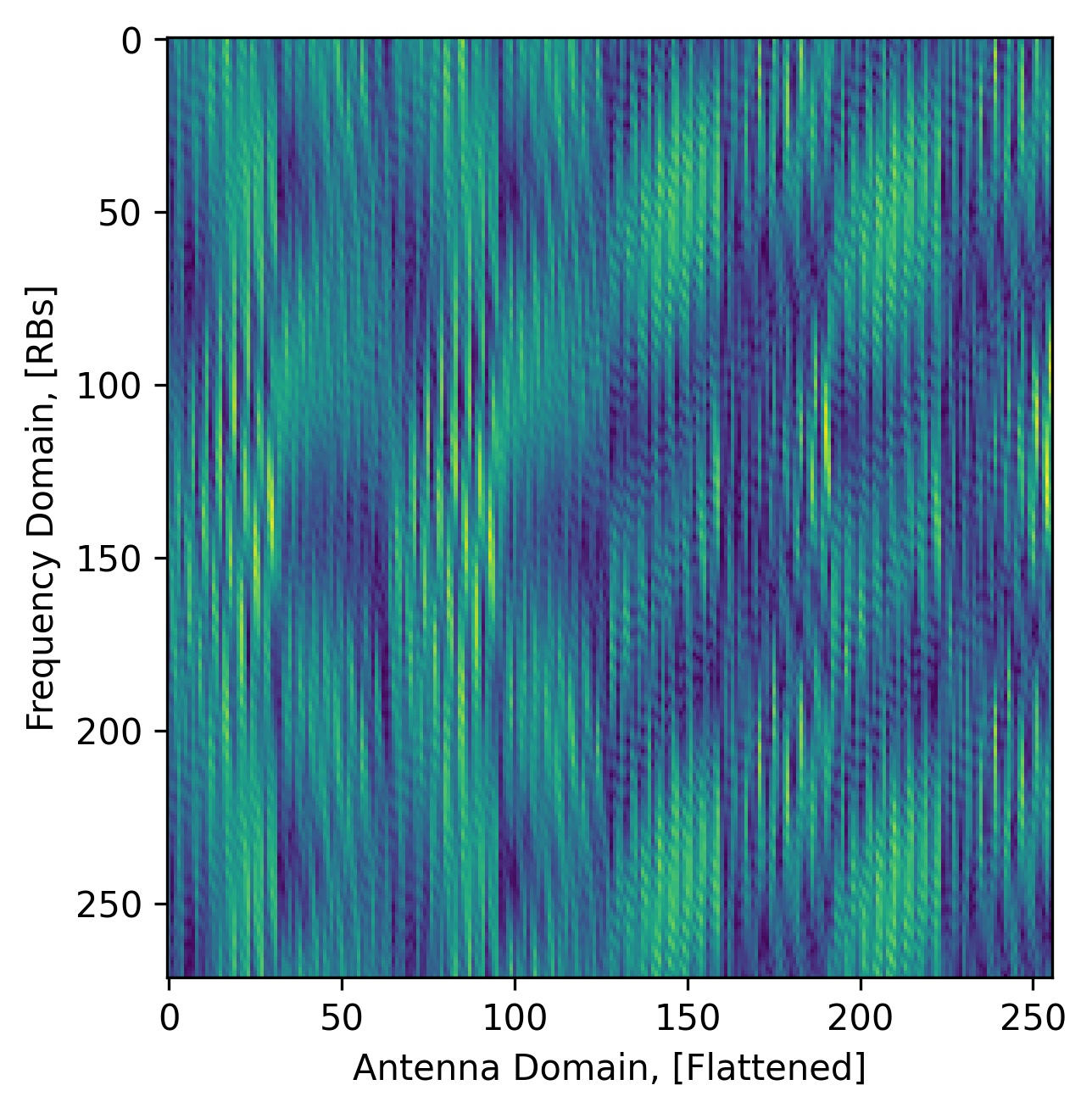}
        \caption{Ground Truth - \textit{High Mobility}}
        \label{fig:perception-distortion-d}
    \end{subfigure}\hfill
    \begin{subfigure}[t]{0.32\linewidth}
        \centering
        \includegraphics[width=\linewidth]{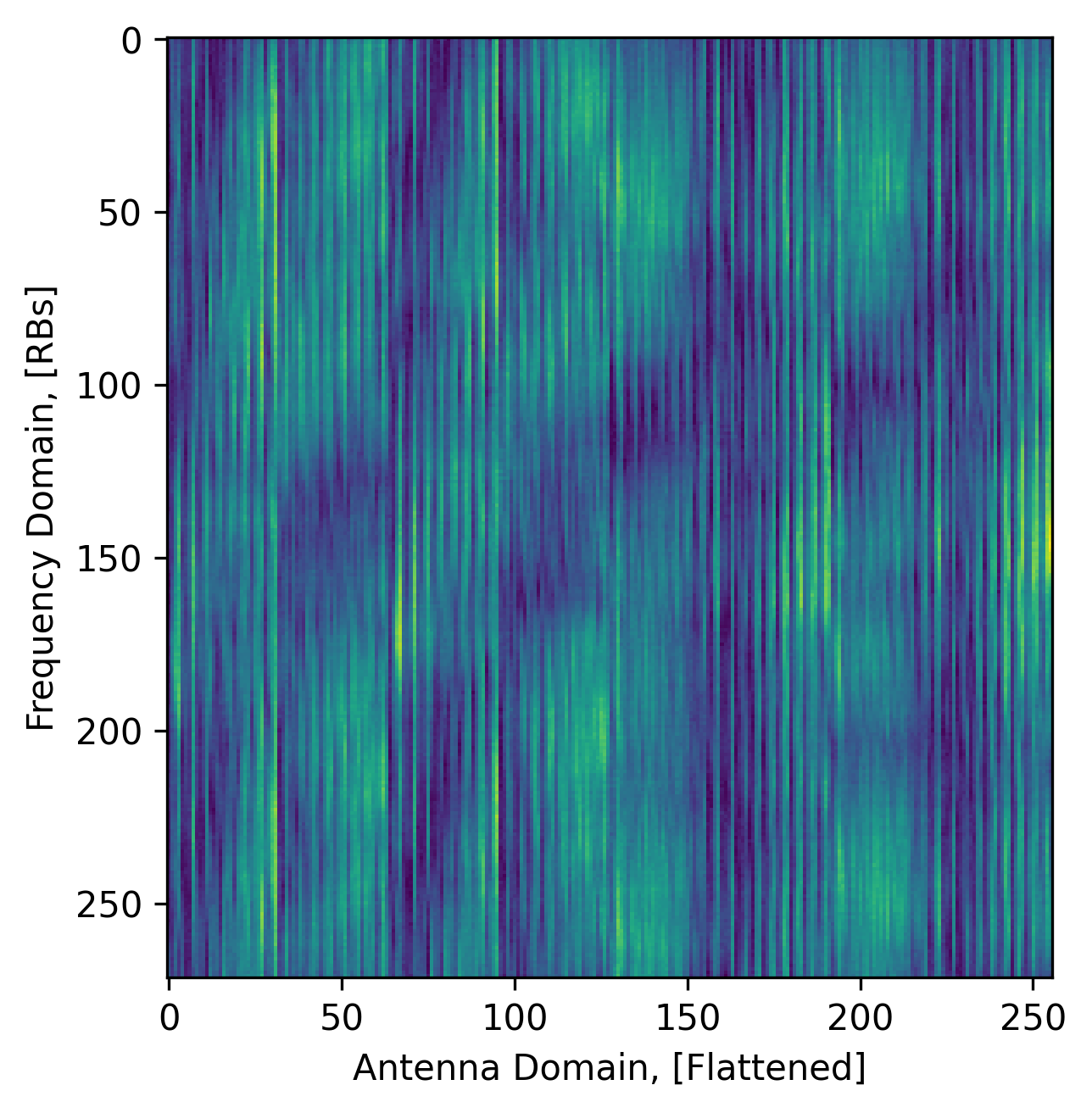}
        \caption{DM Estimate, NMSE=1.96, CS=0.50.}
        \label{fig:perception-distortion-e}
    \end{subfigure} \hfill
    \begin{subfigure}[t]{0.32\linewidth}
        \centering
        \includegraphics[width=\linewidth]{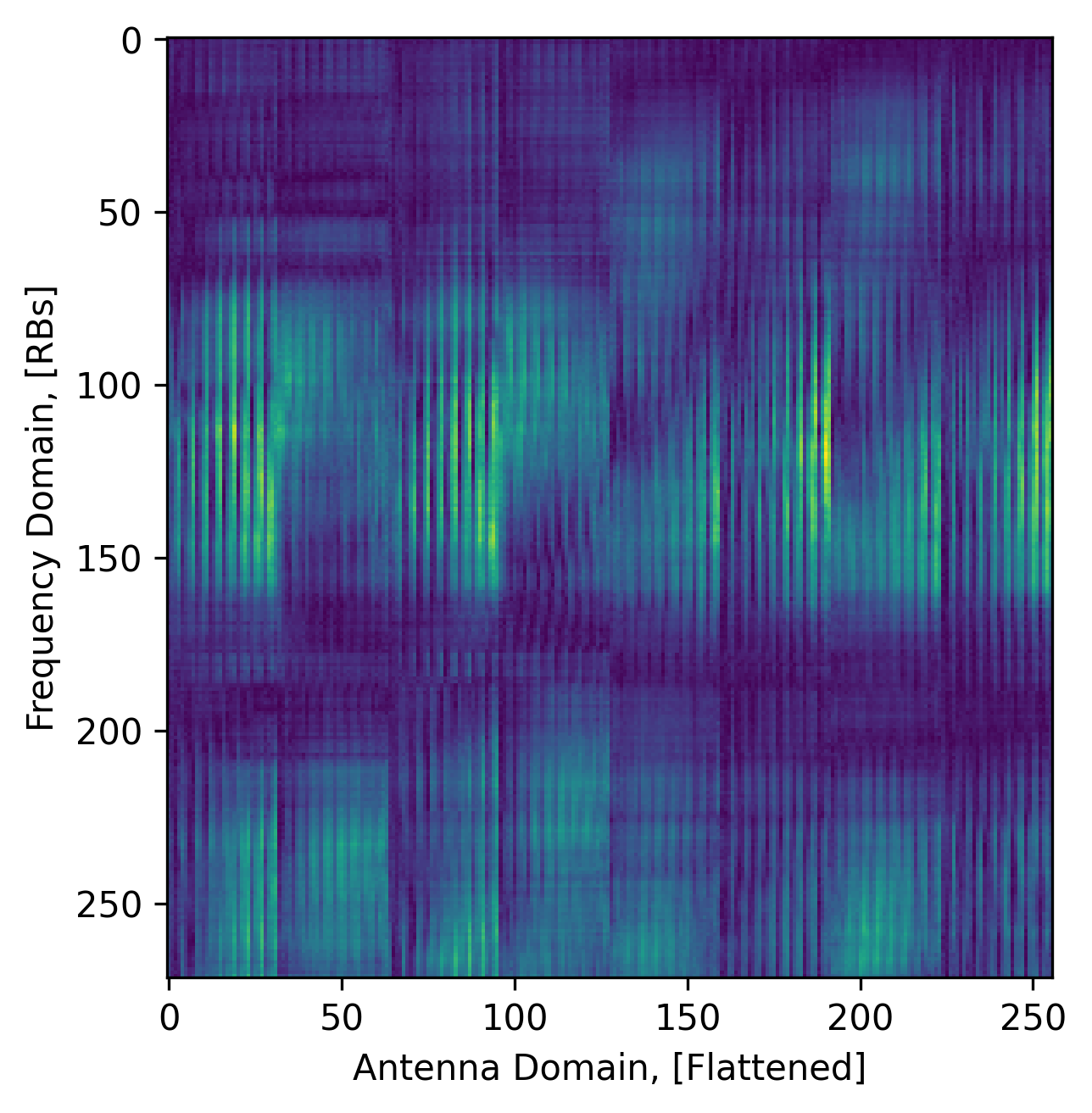}
        \caption{SL Estimate, NMSE=0.79, CS=0.71.}
        \label{fig:perception-distortion-f}
    \end{subfigure}
    \caption{The perception-distortion tradeoff visualized in the context of channel estimation.}
    \label{fig:perception-distortion}
    \vspace{-0.35cm}
\end{figure*}

\begin{figure*}[t]
    \centering
    \begin{subfigure}{0.475\textwidth}
        \centering
        \includegraphics[width=\textwidth]{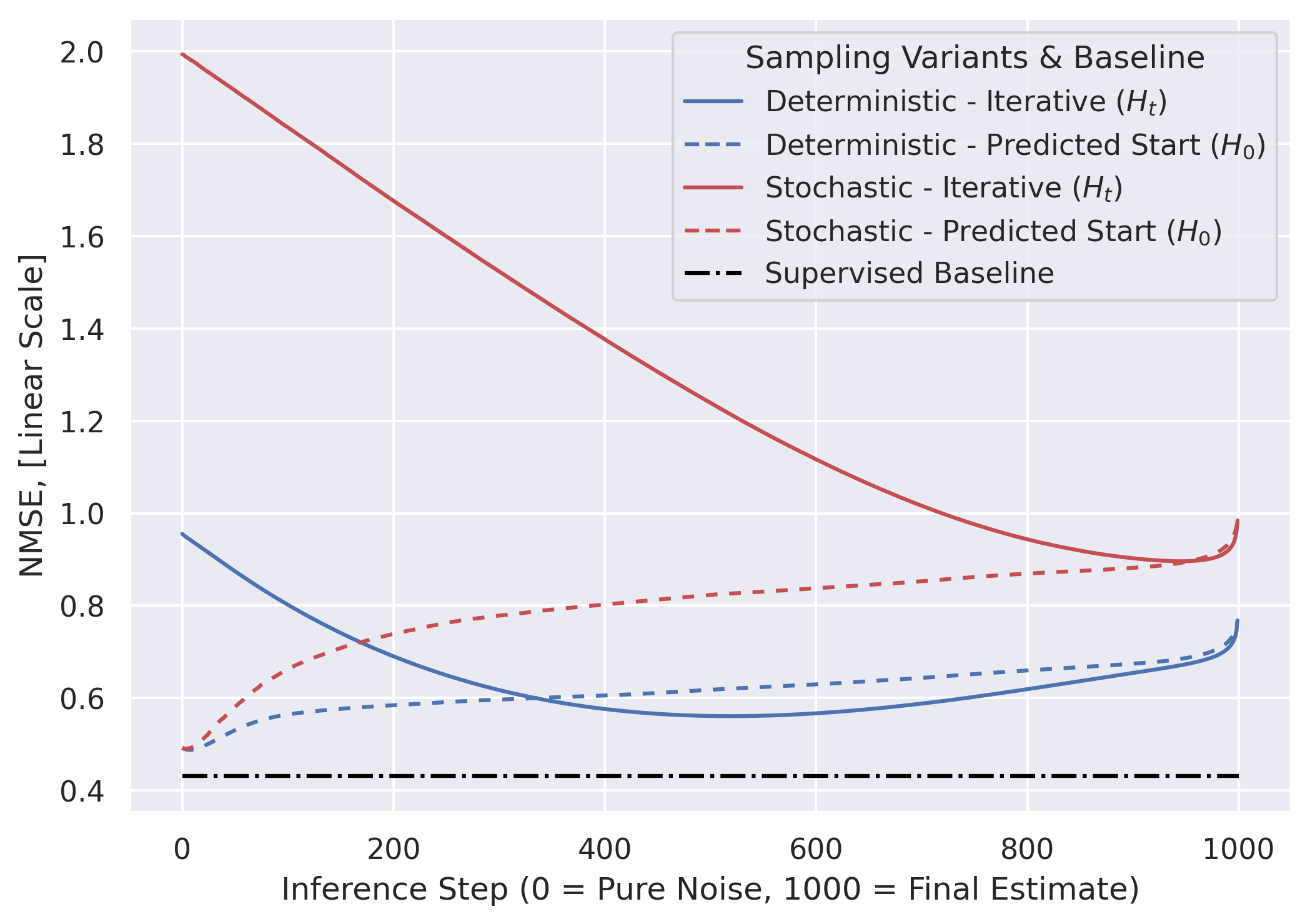}
        \caption{NMSE trajectory for \text{low mobility} scenario.}
        \label{fig:trajectory_low_nmse}
    \end{subfigure}
    \hfill
    \begin{subfigure}{0.475\textwidth}
        \centering
        \includegraphics[width=\textwidth]{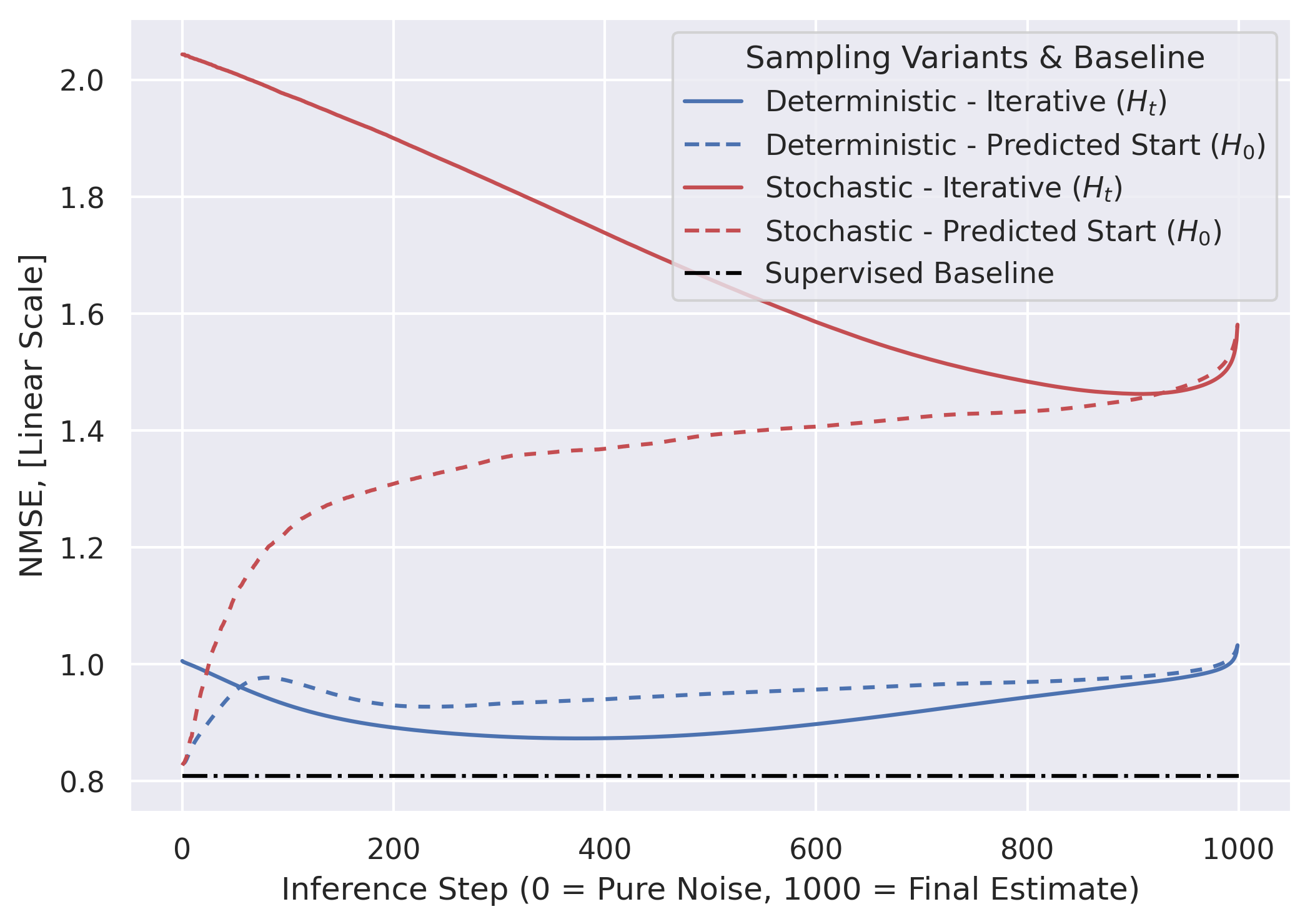}
        \caption{NMSE trajectory for \text{high mobility} scenario.}
        \label{fig:trajectory_high_nmse}
    \end{subfigure} \\ \vspace{0.25cm}
    \begin{subfigure}{0.475\textwidth}
        \centering
        \includegraphics[width=\textwidth]{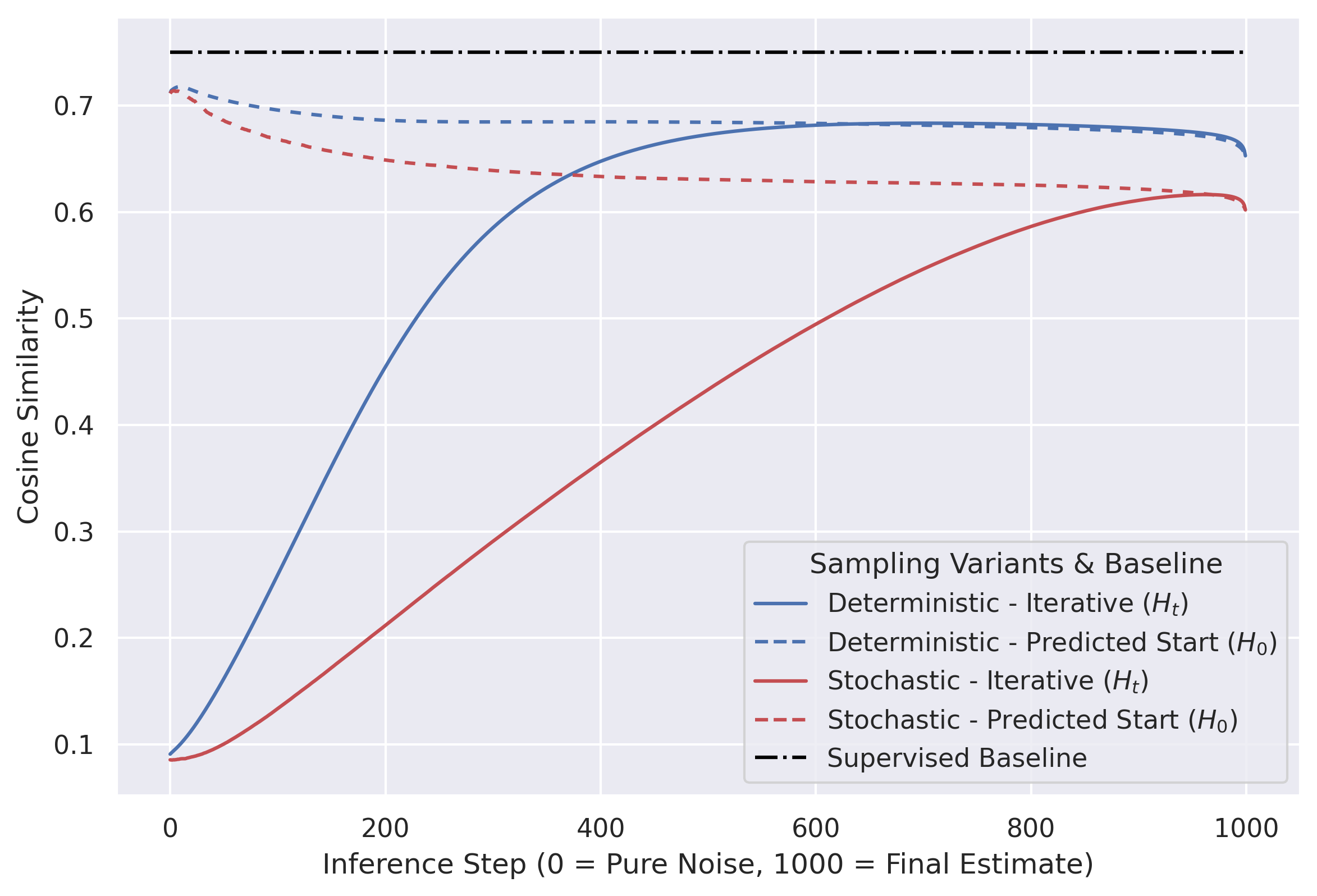}
        \caption{Cosine Similarity trajectory for \text{low mobility} scenario.}
        \label{fig:trajectory_low_cos}
    \end{subfigure}
    \hfill
    \begin{subfigure}{0.475\textwidth}
        \centering
        \includegraphics[width=\textwidth]{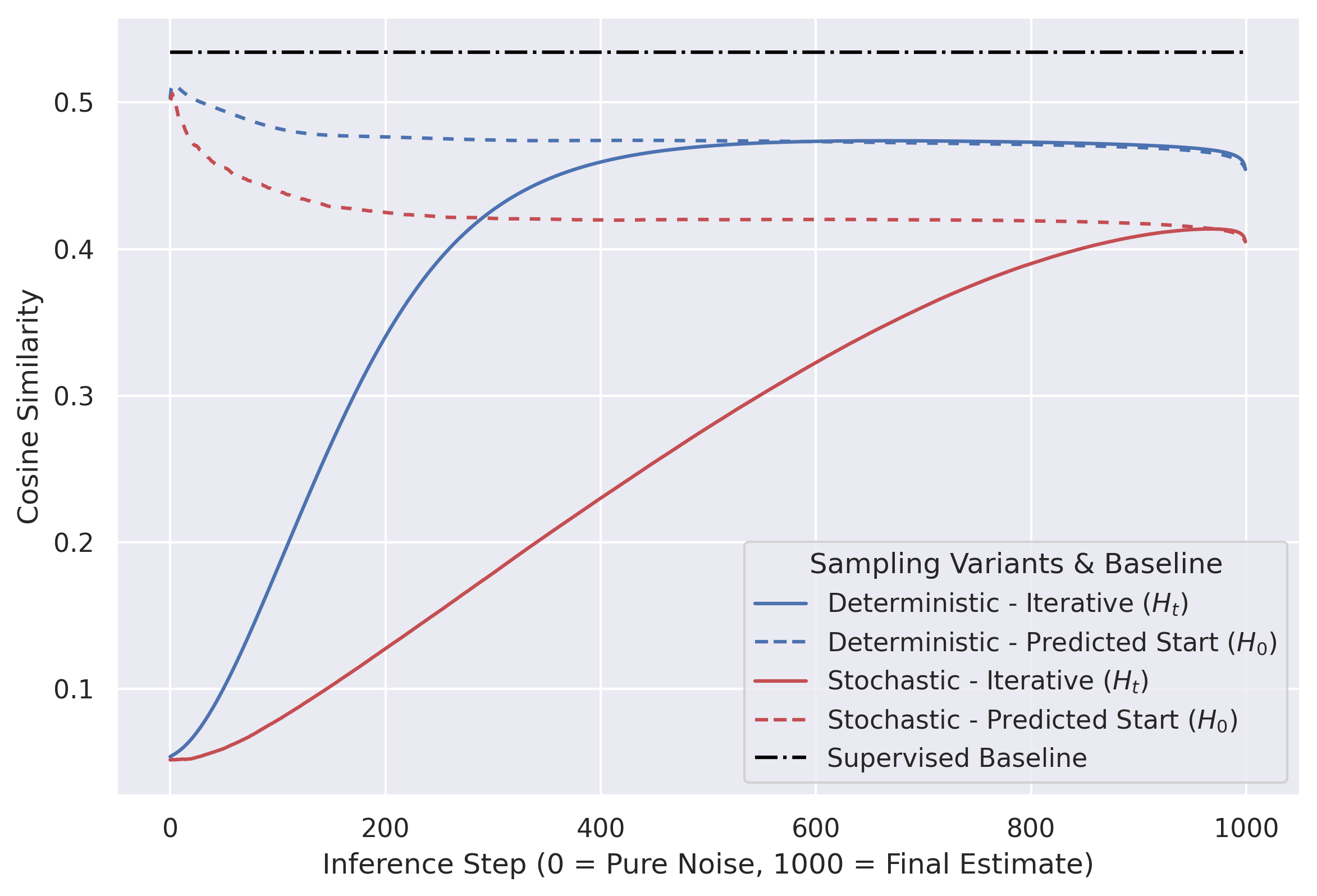}
        \caption{Cosine Similarity trajectory for \text{high mobility} scenario.}
        \label{fig:trajectory_high_cos}
    \end{subfigure}
    \caption{Inference trajectories of the \ac{DM} over the reverse diffusion process. \textit{Stochastic Sampling (Red):} Structural refinement inherently increases distortion, illustrating the perception-distortion tradeoff. \textit{Deterministic Sampling (Blue):} Propagating only the mean limits this degradation, but peak performance remains strictly inferior to the \ac{SL} baseline (black).}
    \label{fig:trajectory_plots}
    \vspace{-0.25cm}
\end{figure*}

This phenomenon is visually evident in the illustrative channel predictions shown in Fig. \ref{fig:perception-distortion}, which compares the ground-truth channels against the \ac{DM} and the discriminative \ac{SL} model estimates. 
In the low mobility scenario (Fig. \ref{fig:perception-distortion}a--c), both models produce seemingly accurate estimates. 
However, closer inspection reveals distinct structural differences. The \ac{DM} successfully captures fine-grained, high-frequency physical details --- particularly noticeable in the left half of Fig. \ref{fig:perception-distortion-b}. 
In contrast, the discriminative \ac{SL} model visibly smooths out these variations. While the generative approach actively maps its estimate to the manifold of physically realistic channels, the discriminative model defaults to a safer, blurred \ac{CME}.
The divergence between the two paradigms is particularly pronounced in the significantly more challenging high mobility scenario (Fig. \ref{fig:perception-distortion}d--f). 
Under severe uncertainty --- which manifests specifically for the resource blocks more affected by channel aging --- the discriminative model predictably falls back to the statistical mean. 
This results in heavily washed-out predictions that approach zero (visualized as dark, featureless regions in Fig. \ref{fig:perception-distortion}f). 
The \ac{DM}, however, avoids collapsing to the mean, successfully generating a complete, structurally coherent channel (Fig. \ref{fig:perception-distortion}e) that maintains the physical textures of the ground truth. 
Despite the \ac{DM}'s clear superiority in generating plausible channels, these fine-grained structural variations are statistically difficult to predict with exact point-wise precision. A minor phase or spatial shift in a highly detailed prediction incurs a \ac{MSE} penalty compared to a safe, conservative mean prediction. This illustrates why a properly trained, physically realistic \ac{DM} should inherently exhibit higher distortion compared to a discriminative baseline specifically trained via \ac{MSE}.

\begin{figure*}[t]
    \centering
    \begin{subfigure}{0.475\textwidth}
        \centering
        \includegraphics[width=\textwidth]{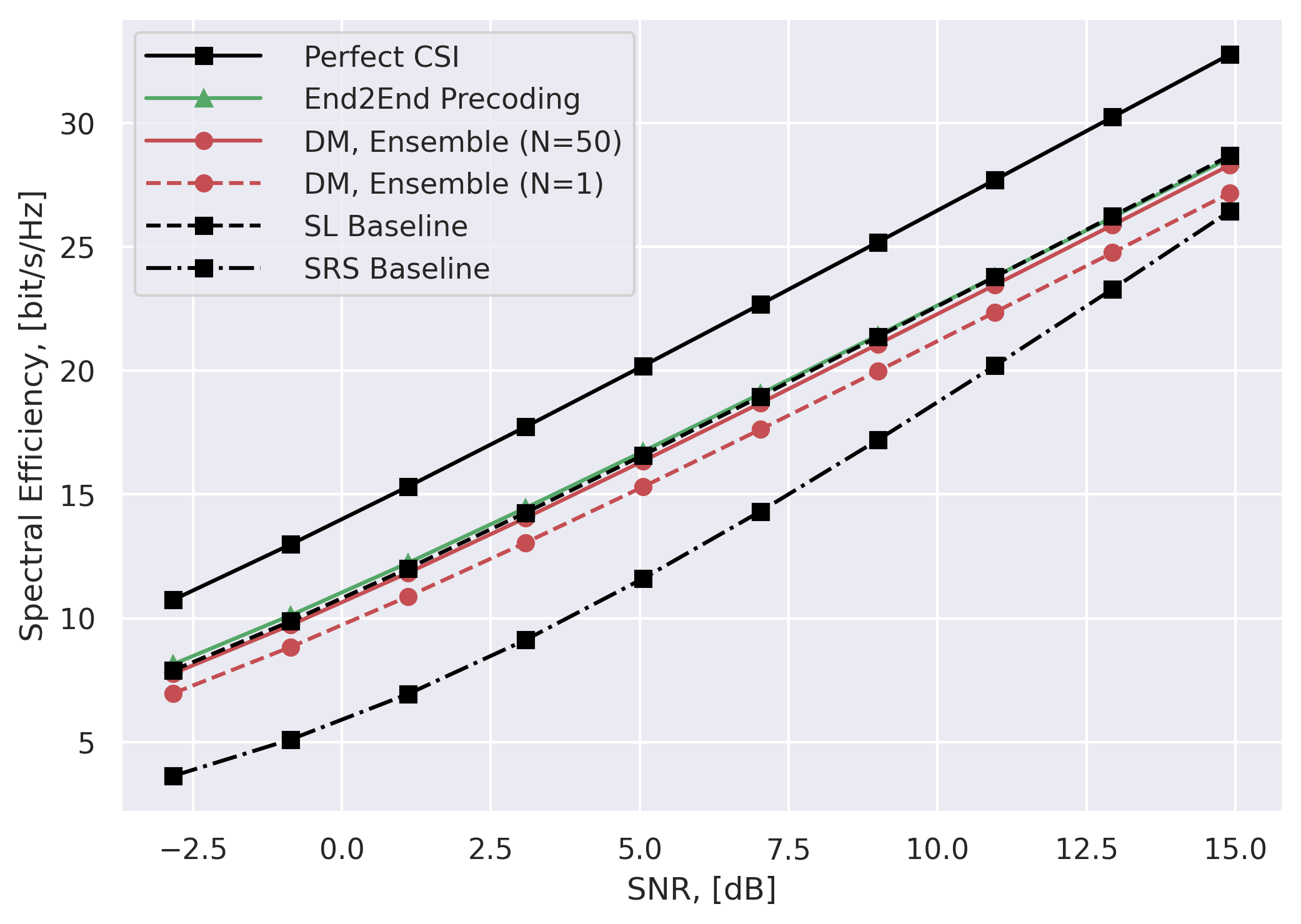}
        \caption{Low mobility.}
        \label{fig:se_performance_low}
    \end{subfigure}
    \hfill
    \begin{subfigure}{0.475\textwidth}
        \centering
        \includegraphics[width=\textwidth]{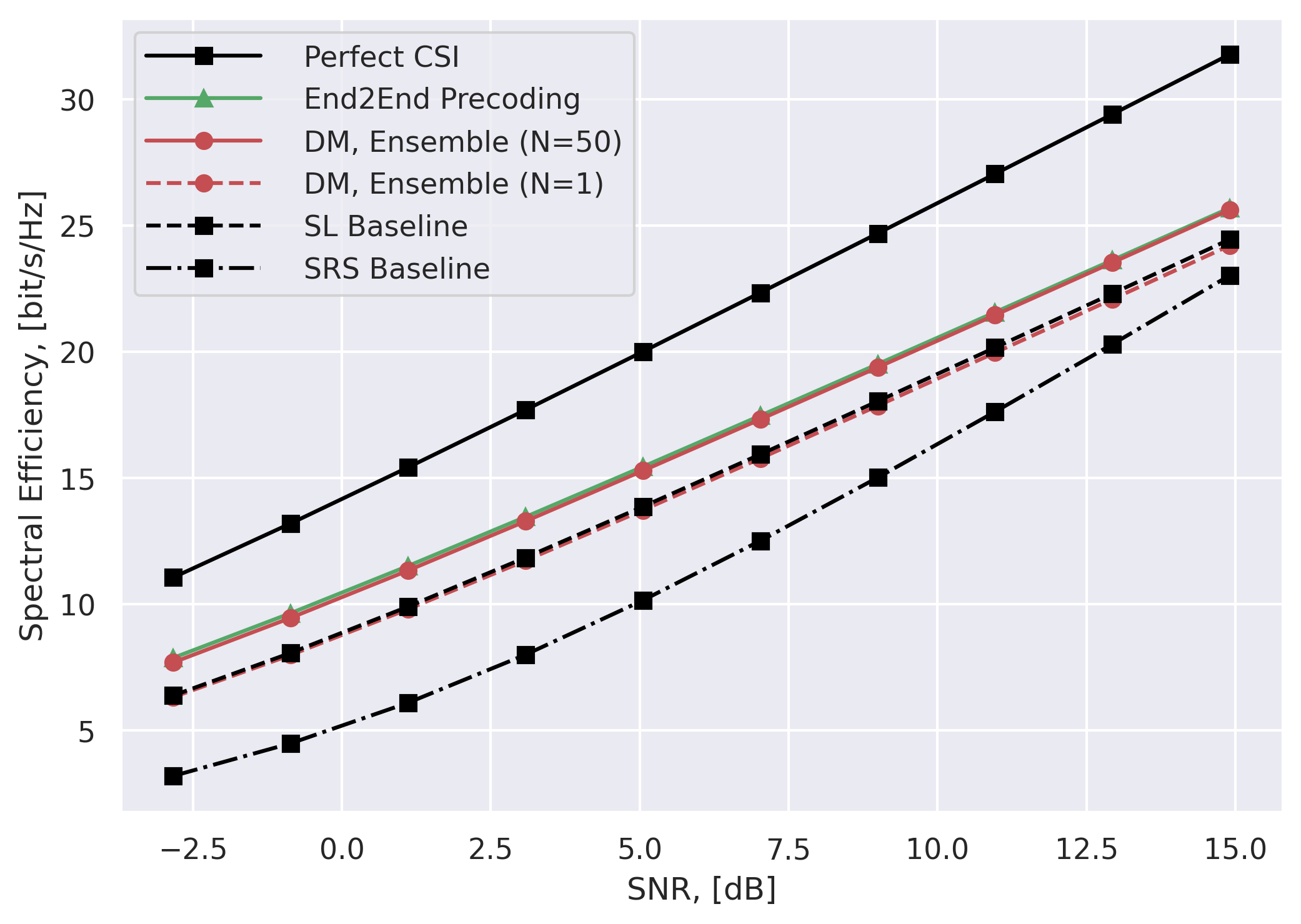}
        \caption{High mobility.}
        \label{fig:se_performance_high}
    \end{subfigure}
    \caption{Downlink \ac{SE} performance under different mobility regimes. \textit{(a) Low Mobility:} Under low structural uncertainty, the conditional posterior is tightly concentrated, rendering the Jensen gap negligible. \textit{(b) High Mobility:} Under severe uncertainty, the deterministic \ac{SL} point-estimate falls victim to the Jensen gap. Fully marginalizing over the posterior via the \ac{DM} ensemble ($N=50$) closes this gap, matching the E2E upper bound without sacrificing modularity.}
    \label{fig:se_performance}
    \vspace{-0.25cm}
\end{figure*}

Despite this suboptimality of generative models in terms of absolute distortion, recent literature frequently reports competitive \ac{NMSE} performance using \acp{DM} (see Section \ref{sec:related_work}). 
This apparent contradiction can be explained by examining the internal dynamics of the inference process. 
As also identified by \cite{fesl2024asymptotic}, a trained \ac{DM} implicitly embeds an \ac{MSE}-optimal estimator within its reverse trajectory. 
Specifically, at the very first step of the reverse process ($t=T$), the model receives pure Gaussian noise that carries zero information about the target channel\footnote{At $t=T$, the ``jump-to-clean'' prediction $\hat{\mathbf{x}}_0 = \frac{1}{\sqrt{\bar{\alpha}_T}}(\mathbf{x}_T - \sqrt{1-\bar{\alpha}_T}\boldsymbol{\epsilon}_\theta(\mathbf{x}_T, T, \mathbf{Y}^{(p)}))$ relies on $\mathbf{x}_T \sim \mathcal{N}(\mathbf{0}, \mathbf{I})$. Because the noise input is uninformative, the objective in \eqref{eq:mean_prediction} forces $\boldsymbol{\epsilon}_\theta$ to minimize the \ac{MSE} solely based on the conditioning $\mathbf{Y}^{(p)}$.}. 
Consequently, the network is forced to predict the underlying clean channel based entirely on the conditioning $\mathbf{Y}^{(p)}$, effectively acting as a \ac{CME}.
This explains why reducing (or totally suppressing) sampling variance, taking fewer steps, or extracting early $\hat{\mathbf{x}}_0$ predictions yields solid \ac{MSE} performance, albeit at the deliberate sacrifice of the perceptual quality the \ac{DM} is inherently designed to achieve.
To empirically evaluate this, we analyze the inference trajectories of our trained \ac{DM} across the diffusion steps of the reverse process. As shown in Fig. \ref{fig:trajectory_plots}, we first examine standard stochastic sampling, which follows the learned transition kernel, and contrast it with a deterministic variant \cite{fesl2024asymptotic}. The deterministic approach\footnote{Fesl et al. \cite{fesl2024asymptotic} formally utilized this deterministic approach for asymptotic denoising, we employ it here as an analytical tool to characterize the strict limits of iterative distortion performance.} suppresses stochasticity by propagating only the conditional mean $\boldsymbol{\mu}_\theta(\mathbf{H}_t, t)$, serving as a mechanism to evaluate the estimate when forced to minimize exploration of the posterior distribution. For both variants, we track the iterative state $\mathbf{H}_{t}$} and the predicted ``jump-to-clean'' start $\hat{\mathbf{H}}_0$.
The resulting trajectories in Fig. \ref{fig:trajectory_plots} provide an empirical characterization of the trade-off between distortion and structural integrity. In the stochastic case (red lines), we observe a severe deterioration in \ac{NMSE} and \ac{GCS} as the inference approaches the final step ($t \to 0$). This occurs because the model ``commits" to a specific, physically plausible channel realization. While this iterative refinement restores sharp structural details (improving perception), it inevitably introduces spatial and phase ``hallucinations'' relative to the exact ground truth, causing a sharp penalty in distortion.
In the deterministic variant (blue lines), the trajectories predictably remain much closer to the statistical mean. Still, the absolute best distortion for this variant is achieved at the very earliest stages of inference (high $t$), where the \ac{DM} objective effectively acts as the aforementioned discriminative \ac{MSE} predictor. However, even the best performance achieved by the \ac{DM} remains inferior to that of the discriminative baseline (dash-dotted line). This performance ceiling highlights a fundamental \textit{capacity gap}: whereas the discriminative \ac{SL} model dedicates its entire parameter budget strictly to direct distortion minimization, the \ac{DM} must distribute its capacity to learn the complex transition score across the entire noise schedule. 
Furthermore, we observe a subsequent decline in the deterministic trajectory over time, which we attribute to the negative impact of accumulated error propagation across the iterative steps. Ultimately, these trajectory dynamics confirm that the computational overhead of iterative diffusion is not intended to minimize \ac{MSE}. Rather, it is designed to navigate the posterior manifold toward physically plausible realizations that a point-estimator cannot represent. Consequently, when designing wireless systems, generative approaches must be carefully benchmarked against classical discriminative models to ensure that the tradeoff between statistical accuracy and distributional fidelity aligns with the specific downstream task.

\subsection{Characterizing Jensen Gap for Precoding}
\label{sec:experiments_precoding}

Despite \acp{SGM}' inherent suboptimality in point-wise distortion, their true utility emerges when considering non-linear downstream tasks of the estimation process such as the \ac{DL} precoding introduced Section \ref{sec:precoding_task}. 
In this section, we evaluate the \ac{DM}-based precoding strategy described in Algorithm \ref{alg:generative_precoding} against a conventional baseline that involves SVD-based precoding derived from the discriminative SL channel estimate $\hat{\mathbf{H}} = \mathbb{E}[\mathbf{H} \mid \mathbf{Y}^{(p)}]$. To observe the effect of Monte Carlo marginalization, we test the \ac{DM} using ensemble sizes of $N=1$ (a single random posterior draw) and $N=50$ samples. To systematically evaluate the capability of the \ac{DM} to offset the excess risk for \ac{DL} precoding, we further compare our generative method against a fully \ac{E2E} trained model\footnote{The \ac{E2E} baseline employs a task-aware objective, extracting the precoder $\mathbf{W}$ from a latent estimate $\hat{\mathbf{H}}$ via a differentiable eigen-decomposition of its Gramian. It is trained end-to-end to maximize the same energy functional as the \ac{DM}, $\max \sum_{k} \|\mathbf{H}_k \mathbf{W}_k\|_F^2$, naturally enforcing $\mathbf{W}_k^H \mathbf{W}_k = \mathbf{I}_{N_L}$.}, acting as an empirical upper bound. All variants are evaluated on their achievable \ac{MIMO} spectral efficiency, assuming equal power allocation across $N_{TX}=4$ spatial layers.

The resulting \ac{SE} performance is characterized in Table \ref{tab:results_se}, with detailed curves over the \ac{SNR} provided in Fig. \ref{fig:se_performance}. For the low mobility scenario in Fig. \ref{fig:se_performance}a, channel aging is minimal, meaning predictive uncertainty is low and the conditional posterior remains tightly concentrated around its mean. Consequently, the excess risk is practically negligible. 
The discriminative \ac{SL} baseline achieves virtually identical \ac{SE} to the E2E and \ac{DM} models, demonstrating that a standard \ac{CME} is entirely sufficient for optimal precoding when uncertainty is strictly bounded\footnote{Note, that with $0.61\%$ the \ac{SL} model achieves a smaller gap than the \ac{DM} with $2.02\%$. We relate this to the capacity gap discussed in Section \ref{sec:experiments_perception}.}. 
However, the limitations of deterministic estimation are exposed in the high mobility scenario in Fig. \ref{fig:se_performance}b. Under severe channel aging, the posterior distribution widens significantly. As dictated by Jensen's inequality, plugging the washed-out conditional mean of the \ac{SL} baseline directly into the non-linear capacity functional noticeably penalizes the achievable rate. As detailed in Table \ref{tab:results_se}, this creates a $8.95\%$ performance penalty relative to the empirical E2E upper bound. 
To illustrate the necessity of proper posterior marginalization, we include the performance of a single generative draw ($N=1$). While this single physically realistic sample avoids the structural smoothing of the \ac{CME}, it fails to capture the full statistical volume of the posterior uncertainty, remaining clearly insufficient to reach the optimal bound (yielding a $9.85\%$ deficit). Crucially, it is only by evaluating the expected posterior Gramian via the larger ensemble ($N=50$) that the generative approach closes the Jensen gap. By fully marginalizing over the uncertainty, the \ac{DM} strictly matches the performance trajectory of the specialized E2E baseline --- shrinking the performance gap to a mere $0.85\%$. 
This validates that generative models can achieve global \ac{E2E} optimality while preserving the modularity and interpretability of classical communication pipelines.

\begin{table}[t!]
    \centering
    \resizebox{\linewidth}{!}{
    \begin{tabular}{@{} l l c c @{}}
        \toprule
        \textbf{Scenario} & \textbf{Model} & \textbf{SE [bps/Hz]} $\uparrow$ & \textbf{$\Delta$ to E2E} $\uparrow$ \\
        \midrule
        \multirow{6}{*}{Low Mobility} 
        & Perfect CSI & $21.556$ & -- \\
        & End-to-End Model & $18.078$ & ref. \\
        & SRS Baseline & $13.786$ & -- \\
        \cmidrule{2-4}
        & SL Model & $\mathbf{17.968}$ & $\mathbf{-0.61}\%$ \\
        & DM Model ($T=50$) & $17.713$ & $-2.02\%$ \\
        & DM Model ($T=1$) & $16.696$ & $-7.65\%$ \\
        \midrule
        \multirow{6}{*}{High Mobility} 
        & Perfect CSI & $21.268$ & -- \\
        & End-to-End Model & $16.583$ & ref. \\
        & SRS Baseline & $12.044$ & -- \\
        \cmidrule{2-4}
        & SL Model & $15.100$ & $-8.95\%$ \\
        & DM Model ($T=50$) & $\mathbf{16.442}$ & $\mathbf{-0.85\%}$ \\
        & DM Model ($T=1$) & $14.949$ & $-9.85\%$ \\
        \bottomrule
    \end{tabular}
    }
    \caption{Downlink \ac{SE}. The gap ($\Delta$) is calculated relative to the empirical upper bound established by the \ac{E2E} model.}
    \label{tab:results_se}
    \vspace{-0.35cm}
\end{table}

\subsection{Link Between Mutual Information and Generative Gain}
\label{sec:experiments_mutual_information}

The contrasting performance between the low and high mobility regimes in Table \ref{tab:results_se} suggests a broader statistical trend: the utility of the generative approach --- and the severity of the Jensen gap --- appears fundamentally linked to the uncertainty of the estimation problem and the informativeness of the pilot observations regarding the true channel.

This notion can be formalized through the concept of mutual information (\ac{MI}), which is defined as the expected Kullback-Leibler (KL) divergence between the posterior and the prior and satisfies the entropy reduction identity:
\begin{align}
\label{eq:mutual_info_def}
I(\mathbf{H}; \mathbf{Y}^{(p)}) &= \mathbb{E}_{\mathbf{Y}^{(p)}} \left[ D_{\mathrm{KL}}\!\left( p(\mathbf{H}\mid \mathbf{Y}^{(p)}) \,\|\, p(\mathbf{H}) \right) \right] \\
&= \mathcal{H}(\mathbf{H}) - \mathbb{E}_{\mathbf{Y}^{(p)}} \left[ \mathcal{H}(\mathbf{H}\mid \mathbf{Y}^{(p)}) \right],
\end{align}
which demonstrates that conditioning on $\mathbf{Y}^{(p)}$ reduces the entropy $\mathcal{H}$ of the channel on average.
This lens clarifies our empirical results. Under high \ac{MI} (e.g., low mobility), the posterior concentrates, and the \ac{CME} remains a structurally accurate channel estimate. 
However, as mobility increases and information drops, the posterior broadens. In such low-information regimes --- approaching the zero-information limit of pilot-free precoding --- deterministic estimators must average over diverging hypotheses to minimize \ac{MSE}. {This smoothing distorts the spatial structure of the channel, leading to a significant reduction in spectral efficiency.}

Interestingly, \acp{DM} provide a way to artificially control the precision of the information flow during inference. This lets us study such relationship without being tied to the actual physical mobility of the channel. This is achieved through \ac{CFG}, where the guidance scale $\lambda \in [0,1]$ serves as a controllable dial for the conditioning signal. As described in \eqref{eq:cfg}, setting $\lambda=1$ recovers standard conditional sampling, exploiting all available \ac{MI}. Conversely, setting $\lambda=0$ completely severs the conditioning, effectively simulating the extreme zero-information limit ($I(\mathbf{H};\mathbf{Y}^{(p)})=0$) by forcing the model to sample strictly from the unconditional prior $p(\mathbf{H})$.

\begin{figure}\centering
    \includegraphics[width= 0.95\linewidth]{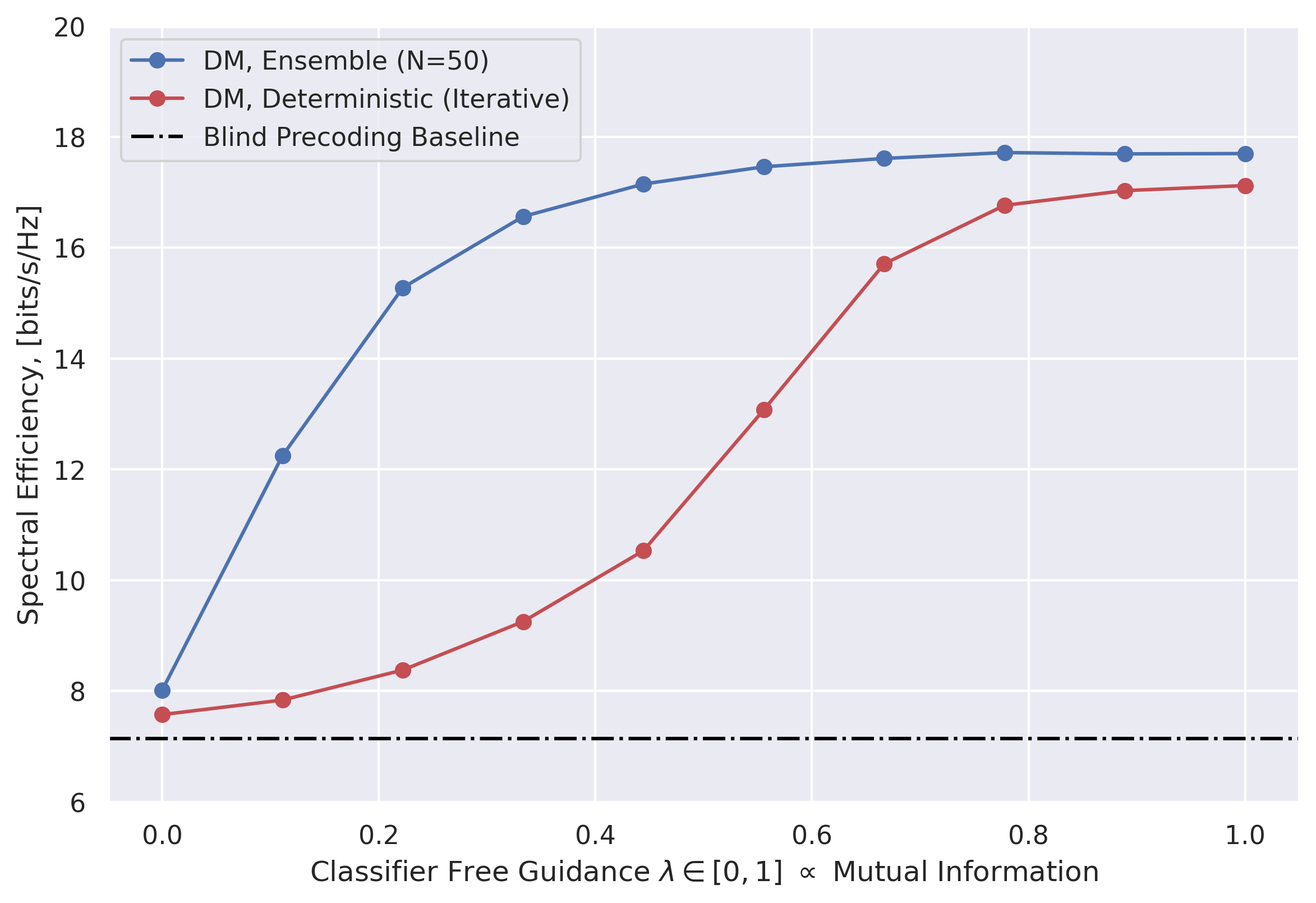}
    \caption{The Jensen gap evaluated over a mutual information proxy (\ac{CFG} $\lambda$) for the \textit{low mobility} case. As the conditioning becomes less informative ($\lambda < 1$), the deterministic estimate collapses, massively widening the performance gap compared to the properly marginalized \ac{DM} ensemble.}
    \label{fig:cfg_plot}
\end{figure}

Fig. \ref{fig:cfg_plot} tracks the resulting \ac{SE} across this information continuum, comparing the properly marginalized \ac{DM} ensemble ($N=50$) against a purely deterministic \ac{DM} 
estimate\footnote{Via \ac{CFG}, the modified noise estimate is given by $\hat{\boldsymbol{\epsilon}} = \boldsymbol{\epsilon}_\theta(\mathbf{H}_t, t, \emptyset) + \lambda \big( \boldsymbol{\epsilon}_\theta(\mathbf{H}_t, t, \mathbf{Y}^{(p)}) - \boldsymbol{\epsilon}_\theta(\mathbf{H}_t, t, \emptyset) \big)$. Analogous to the evaluation in Section \ref{sec:perception_distortion_tradeoff}, we again build the deterministic estimator by propagating only the mean}.
At full information ($\lambda=1$), the {performance} gap between the two approaches is relatively small\footnote{Note that a baseline gap at $\lambda=1$ remains here because the deterministic \ac{DM} exhibits a slightly worse \ac{NMSE} than a dedicated discriminative model trained via MSE loss, due to the model capacity and error propagation effects discussed in Section \ref{sec:experiments_perception}}.
However, as $\lambda$ decreases --- simulating a progressively less informative conditioning signal --- the smoothing of the conditional mean rapidly degrades the deterministic estimate. Its performance plummets significantly earlier than the Bayesian precoder, causing the Jensen gap to expand drastically in the mid-to-low information regime. 
By contrast, the generative \ac{DM} sustains higher \ac{SE} by preserving structural diversity through proper statistical sampling. As $\lambda \to 0$, the gap inherently narrows. At this extreme unconditional limit, both models converge toward the theoretical minimum dictated by the baseline spatial entropy $\mathcal{H}(\mathbf{H})$ of the UMa dataset, aligning with the pilotless precoding case --- given by uniform power allocation across all antennas, with zero directional information. 
Ultimately, this parameter sweep confirms our core hypothesis: the computational overhead of generative marginalization yields its highest dividends outside of high-information environments, where deterministic point-estimators inherently suffer structural collapse.

\section{Conclusion}
\label{sec:conclusion}
In this work, we have reframed the role of \acp{DM} in wireless communications, moving beyond the pursuit of \ac{MSE} optimality. Our results demonstrate that while generative models are inherently penalized by the perception-distortion tradeoff in terms of point-wise accuracy, their true value lies in their ability to resolve structural uncertainty. By providing access to the full channel posterior, \acp{DM} serve as general-purpose engines for channel-aware decision-making, capable of optimizing non-linear functionals that are mathematically out of reach for deterministic estimators. Crucially, our findings show that generative marginalization bridges the gap between the global optimality of \ac{E2E} learning and the practical necessity of modular communication pipelines. This "train once, estimate many" paradigm allows for the deployment of a single generative channel estimation module that can be used for various downstream tasks such as \ac{SRS}-based precoding --- without the rigid task-dependency of \ac{E2E} models. While the current computational overhead of iterative diffusion remains a challenge for real-time chip implementation, this work validates the generative framework as a powerful new alternative to traditional estimation. To translate these generative gains into next-generation hardware, future work must prioritize efficiency improvements such as distilled sampling. Furthermore, extending this framework to several downstream tasks like decoding, equalization, and detection will be key, requiring new optimization strategies specifically tailored to these diverse downstream objectives.

\begin{appendices}

\section{End-to-End Learning vs. Posterior Marginalization}
 An E2E model directly maximizes the global task functional $\mathcal{T}(\mathbf{x}, \mathbf{w})$. By the law of total expectation, and under the assumption of unlimited model capacity, this global objective decomposes exactly into the expected posterior utility:
\begin{equation}
\begin{aligned} 
    & \max_{\mathbf{w}=g(\mathbf{y};\theta)} \mathbb{E}_{\mathbf{x},\mathbf{y}}\big[\mathcal{T}(\mathbf{x}, \mathbf{w})\big] \\
    &= \mathbb{E}_{\mathbf{y}} \left[ \max_{\mathbf{w}} \mathbb{E}_{\mathbf{x}|\mathbf{y}}[\mathcal{T}(\mathbf{x}, \mathbf{w})] \right] \\ &= \int_{\mathcal{Y}} \underbrace{\left( \max_{\mathbf{w}} \int_{\mathcal{X}} \mathcal{T}(\mathbf{x}, \mathbf{w}) p(\mathbf{x}\mid\mathbf{y}) d\mathbf{x} \right)}_{\text{Posterior Expected Utility}} p(\mathbf{y}) d\mathbf{y}.
\end{aligned}
\end{equation}
This reveals that an E2E network implicitly solves for the posterior expected utility at every observation $\mathbf{y}$, yet its learned mapping remains rigidly bound to the specific choice of $\mathcal{T}$.

In contrast, the generative approach explicitly decouples estimation from optimization. By learning the target distribution $p(\mathbf{x}\mid\mathbf{y}; \theta^*) \approx p(\mathbf{x}\mid\mathbf{y})$ independently of the downstream objective, the model acts as a general-purpose estimator. During inference, the optimal action $\mathbf{w}^*(\mathbf{y})$ for \textit{any} arbitrary task is found by explicitly evaluating the identical inner objective:
\begin{equation}
    \mathbf{w}^*(\mathbf{y}) = \arg\max_{\mathbf{w}} \int_{\mathcal{X}} \mathcal{T}(\mathbf{x}, \mathbf{w}) p(\mathbf{x}\mid\mathbf{y}; \theta^*) d\mathbf{x}.
\end{equation}
Therefore, provided the true posterior is accurately learned and the downstream optimization over $\mathbf{w}$ is tractable, posterior marginalization theoretically achieves the exact global optimality of E2E training while strictly preserving the modularity of classical communication pipelines.

\section{Model Architecture and Training Procedure}
\label{app:model_description}

Both the \ac{DM} and the discriminative model share an identical PyTorch transformer encoder/decoder structure, which closely follows the original architecture proposed in \cite{vaswani2017attention}. 
As detailed in Table \ref{tab:model_architecture}, the tokens are formed by patching the input channel into real-valued $128$-dimensional tensors, which are then linearly projected into a $512$-dimensional embedding. 
The discriminative baseline is trained via \ac{ERM} with an \ac{MSE} loss, while the \ac{DM} minimizes a score-matching objective with a cosine noise-schedule to approximate the full posterior distribution.

\begin{table}\centering
    \resizebox{0.95\linewidth}{!}{
    \begin{tabular}{@{} l l @{}}
        \toprule
        \textbf{Module} & \textbf{Configuration} \\
        \midrule
        \multicolumn{2}{l}{\textit{Shared Transformer Backbone}} \\
        Input Patching & $128$ ($64 \times 2$ real/imag. \ac{BS} antennas) \\
        Embedding & Linear Projection $\mathbb{R}^{128} \rightarrow \mathbb{R}^{512}$ \\
        Positional Encoding & Learnable additive encoding \\
        Transformer Encoder & $3$ Layers, $n_{\text{head}}=8$, $d_{\text{ff}}=1024$ \\
        Transformer Decoder & $3$ Layers, $n_{\text{head}}=8$, $d_{\text{ff}}=1024$ \\
        Attention / Dropout & Multi-Head (Self/Cross) / $0.1$ \\
        Output Head & Linear Projection $\mathbb{R}^{512} \rightarrow \mathbb{R}^{128}$ \\
        \midrule
        \multicolumn{2}{l}{\textit{Discriminative Baseline (\ac{SL})}} \\
        Learning Framework & \ac{ERM} \\
        Loss Function & \ac{MSE} \\
        \midrule
        \multicolumn{2}{l}{\textit{Generative Model (\ac{DM})}} \\
        Objective & Score Matching (predicting $\mathbf{H}_0$) \\
        Noise Schedule & Cosine Schedule ($T=1000$ steps) \\
        Inference & Iterative Denoising (fixed encoder) \\
        \midrule
        \multicolumn{2}{l}{\textit{Training \& Dataset Details}} \\
        Optimizer / Learning Rate & Adam / $10^{-4}$ \\
        Training Dataset Size & 100k Samples \\
        Test Dataset Size & 5k Samples \\
        \bottomrule
    \end{tabular}
    }
    \caption{Model Architecture and Training Parameters}
    \label{tab:model_architecture}
    \vspace{-0.5cm}
\end{table}

\section{Optimal Precoding for Expected Received Energy}
\label{app:rayleigh_ritz}

Let $\mathbf{W}_k \in \mathbb{C}^{N_{\text{tx}} \times N_L}$ denote the precoding matrix for subcarrier $k$, constrained to the Stiefel manifold such that $\mathbf{W}_k^H \mathbf{W}_k = \mathbf{I}_{N_L}$. To maximize the expected received energy under the posterior channel distribution $p(\mathbf{H} \mid \mathbf{Y}^{(p)})$, we formulate the problem as the maximization of $\mathbb{E} \big[\|\mathbf{H}_k \mathbf{W}_k\|_F^2\big]$. Using the definition of the Frobenius norm, we can commute the expectation and trace operators as follows:
\begin{align}
    \mathbb{E}_{\mathbf{H} \mid \mathbf{Y}^{(p)}} \Big[ \|\mathbf{H}_k \mathbf{W}_k\|_F^2 \Big] 
    &= \mathbb{E}_{\mathbf{H} \mid \mathbf{Y}^{(p)}} \Big[ \text{tr} \left( \mathbf{W}_k^H \mathbf{H}_k^H \mathbf{H}_k \mathbf{W}_k \right) \Big] \nonumber \\
    &= \text{tr} \left( \mathbf{W}_k^H \, \mathbb{E}_{\mathbf{H} \mid \mathbf{Y}^{(p)}} \big[ \mathbf{H}_k^H \mathbf{H}_k \big] \, \mathbf{W}_k \right) \nonumber \\
    &= \text{tr} \left( \mathbf{W}_k^H \mathbf{R}_k \mathbf{W}_k \right),
\end{align}
where $\mathbf{R}_k \triangleq \mathbb{E}_{\mathbf{H} \mid \mathbf{Y}^{(p)}} [\mathbf{H}_k^H \mathbf{H}_k]$ is the expected posterior Gramian. The optimization problem thus simplifies to:
\begin{equation}
    \max_{\mathbf{W}_k} \text{tr}(\mathbf{W}_k^H \mathbf{R}_k \mathbf{W}_k) \quad \text{subject to} \quad \mathbf{W}_k^H \mathbf{W}_k = \mathbf{I}_{N_L}.
\end{equation}
Maximizing the trace of a projected Hermitian matrix under orthonormal constraints is a standard result in matrix analysis \cite[Sec. 4.3]{horn2012matrix}. It is well established that the maximum is achieved when the column space of $\mathbf{W}_k$ is spanned by the principal eigenvectors of $\mathbf{R}_k$. Consequently, the optimal Bayesian precoder $\mathbf{W}_k^*$ is constructed directly from the $N_L$ dominant eigenvectors. Notably, in the case of a single posterior sample, where the expectation simplifies to the instantaneous channel estimate, this solution is mathematically equivalent to the standard SVD-based precoder.
\end{appendices}

\begingroup
\small
\bibliographystyle{IEEEtran}
\bibliography{references}

@article{ha2025bayesian,
  title={Bayesian Radio Map Estimation: Fundamentals and Implementation via Diffusion Models},
  author={Ha, Tien Ngoc and Romero, Daniel},
  journal={arXiv preprint arXiv:2508.06037},
  year={2025}
}

@article{van2024generative,
  title={Generative AI for physical layer communications: A survey},
  author={Van Huynh, Nguyen and Wang, Jiacheng and Du, Hongyang and Hoang, Dinh Thai and Niyato, Dusit and Nguyen, Diep N and Kim, Dong In and Letaief, Khaled B},
  journal={IEEE Transactions on Cognitive Communications and Networking},
  volume={10},
  number={3},
  pages={706--728},
  year={2024},
  publisher={IEEE}
}

@inproceedings{strasser2025enhancements,
  title={Enhancements in Score-Based Channel Estimation for Real-Time Wireless Systems},
  author={Strasser, Florian and B{\"a}ro, Marion and Utschick, Wolfgang},
  booktitle={2025 28th International Workshop on Smart Antennas (WSA)},
  pages={140--146},
  year={2025},
  organization={IEEE}
}

@misc{fabian2024diracdiffusiondenoisingincrementalreconstruction,
      title={DiracDiffusion: Denoising and Incremental Reconstruction with Assured Data-Consistency}, 
      author={Zalan Fabian and Berk Tinaz and Mahdi Soltanolkotabi},
      year={2024},
      eprint={2303.14353},
      archivePrefix={arXiv},
      primaryClass={eess.IV},
      url={https://arxiv.org/abs/2303.14353}, 
}

@article{fesl2024asymptotic,
  title={On the asymptotic mean square error optimality of diffusion models},
  author={Fesl, Benedikt and B{\"o}ck, Benedikt and Strasser, Florian and Baur, Michael and Joham, Michael and Utschick, Wolfgang},
  journal={arXiv preprint arXiv:2403.02957},
  year={2024}
}

@article{hornik1989multilayer,
  title={Multilayer feedforward networks are universal approximators},
  author={Hornik, Kurt and Stinchcombe, Maxwell and White, Halbert},
  journal={Neural networks},
  volume={2},
  number={5},
  pages={359--366},
  year={1989},
  publisher={Elsevier}
}

@book{horn2012matrix,
  title={Matrix Analysis},
  author={Horn, Roger A and Johnson, Charles R},
  year={2012},
  edition={2nd},
  publisher={Cambridge University Press}
}

@inproceedings{wang2025traversing,
  title={Traversing Distortion-Perception Tradeoff using a Single Score-Based Generative Model},
  author={Wang, Yuhan and Bi, Suzhi and Zhang, Ying-Jun Angela and Yuan, Xiaojun},
  booktitle={Proceedings of the Computer Vision and Pattern Recognition Conference},
  pages={2377--2386},
  year={2025}
}

@book{CoverThomas2006,
  author    = {Cover, Thomas M. and Thomas, Joy A.},
  title     = {Elements of Information Theory},
  edition   = {2nd},
  publisher = {Wiley-Interscience},
  address   = {Hoboken, NJ, USA},
  year      = {2006},
  isbn      = {978-0471241959}
}

@INPROCEEDINGS{10619829,
  author={Luo, Xuanhao and Zhizhen, L. and Peng, Zhiyuan and Dongkuan, X. and Liu, Yuchen},
  booktitle={2024 IFIP Networking Conference (IFIP Networking)}, 
  title={RM-Gen: Conditional Diffusion Model-Based Radio Map Generation for Wireless Networks}, 
  year={2024},
  volume={},
  number={},
  pages={543-548},
  doi={10.23919/IFIPNetworking62109.2024.10619829}
}

@article{vaswani2017attention,
  title={Attention is all you need},
  author={Vaswani, Ashish and Shazeer, Noam and Parmar, Niki and Uszkoreit, Jakob and Jones, Llion and Gomez, Aidan N and Kaiser, {\L}ukasz and Polosukhin, Illia},
  journal={Advances in neural information processing systems},
  volume={30},
  year={2017}
}

@ARTICLE{vincent2011connection,
  author={Vincent, Pascal},
  journal={Neural Computation}, 
  title={A Connection Between Score Matching and Denoising Autoencoders}, 
  year={2011},
  volume={23},
  number={7},
  pages={1661-1674},
  doi={10.1162/NECO_a_00142}}

@article{song2019generative,
  title={Generative modeling by estimating gradients of the data distribution},
  author={Song, Yang and Ermon, Stefano},
  journal={Advances in neural information processing systems},
  volume={32},
  year={2019}
}

@inproceedings{wang2025radiodiff,
  title={RadioDiff-Turbo: Lightweight Generative Large Electromagnetic Model for Wireless Digital Twin Construction},
  author={Wang, Xiucheng and Zheng, Peilin and Cheng, Nan and Sun, Ruijin and Chen, Junting and Tao, Keda and Yin, Zhisheng and Liu, Zhiquan and Zeng, Yong},
  booktitle={IEEE INFOCOM 2025-IEEE Conference on Computer Communications Workshops (INFOCOM WKSHPS)},
  pages={1--6},
  year={2025},
  organization={IEEE}
}

@article{arvinte2022mimo,
  title={MIMO channel estimation using score-based generative models},
  author={Arvinte, Marius and Tamir, Jonathan I},
  journal={IEEE Transactions on Wireless Communications},
  volume={22},
  number={6},
  pages={3698--3713},
  year={2022},
  publisher={IEEE}
}

@ARTICLE{11391481,
  author={Zhang, Xinren and Yu, Jiadong},
  journal={IEEE Transactions on Wireless Communications}, 
  title={Improve the Training Efficiency of DRL for Wireless Communication Resource Allocation: The Role of Generative Diffusion Models}, 
  year={2026},
  volume={25},
  number={},
  pages={11593-11608},
  doi={10.1109/TWC.2026.3660250}}

@article{liang2025diffusion,
  title={Diffusion models as network optimizers: Explorations and analysis},
  author={Liang, Ruihuai and Yang, Bo and Chen, Pengyu and Li, Xianjin and Xue, Yifan and Yu, Zhiwen and Cao, Xuelin and Zhang, Yan and Debbah, M{\'e}rouane and Poor, H Vincent and others},
  journal={IEEE Internet of Things Journal},
  year={2025},
  publisher={IEEE}
}

@book{absil2008optimization,
  title={Optimization algorithms on matrix manifolds},
  author={Absil, P-A and Mahony, Robert and Sepulchre, Rodolphe},
  year={2008},
  publisher={Princeton University Press}
}

@article{hoydis2022sionna,
  title={Sionna: An open-source library for next-generation physical layer research},
  author={Hoydis, Jakob and Cammerer, Sebastian and Aoudia, Fay{\c{c}}al Ait and Vem, Avinash and Binder, Nikolaus and Marcus, Guillermo and Keller, Alexander},
  journal={arXiv preprint arXiv:2203.11854},
  year={2022}
}

@inproceedings{blau2018perception,
  title={The perception-distortion tradeoff},
  author={Blau, Yochai and Michaeli, Tomer},
  booktitle={Proceedings of the IEEE conference on computer vision and pattern recognition},
  pages={6228--6237},
  year={2018}
}

@techreport{huawei2026csi,
  author      = {{Huawei} and {HiSilicon}},
  title       = {Discussion on inference, data collection, and monitoring related aspects for {CSI} compression},
  type        = {3GPP TSG-RAN WG1 Meeting \#124},
  number      = {R1-2600079},
  institution = {3GPP},
  address     = {Gothenburg, Sweden},
  month       = feb,
  year        = {2026}
}

@article{graikos2022diffusion,
  title={Diffusion models as plug-and-play priors},
  author={Graikos, Alexandros and Malkin, Nikolay and Jojic, Nebojsa and Samaras, Dimitris},
  journal={Advances in Neural Information Processing Systems},
  volume={35},
  pages={14715--14728},
  year={2022}
}

@inproceedings{nichol2021improved,
  title={Improved denoising diffusion probabilistic models},
  author={Nichol, Alexander Quinn and Dhariwal, Prafulla},
  booktitle={International conference on machine learning},
  pages={8162--8171},
  year={2021},
  organization={PMLR}
}

@article{song2020score,
  title={Score-based generative modeling through stochastic differential equations},
  author={Song, Yang and Sohl-Dickstein, Jascha and Kingma, Diederik P and Kumar, Abhishek and Ermon, Stefano and Poole, Ben},
  journal={arXiv preprint arXiv:2011.13456},
  year={2020}
}

@article{ho2020denoising,
  title={Denoising diffusion probabilistic models},
  author={Ho, Jonathan and Jain, Ajay and Abbeel, Pieter},
  journal={Advances in neural information processing systems},
  volume={33},
  pages={6840--6851},
  year={2020}
}

@InProceedings{pmlr-v37-sohl-dickstein15,
  title = 	 {Deep Unsupervised Learning using Nonequilibrium Thermodynamics},
  author = 	 {Sohl-Dickstein, Jascha and Weiss, Eric and Maheswaranathan, Niru and Ganguli, Surya},
  booktitle = 	 {Proceedings of the 32nd International Conference on Machine Learning},
  pages = 	 {2256--2265},
  year = 	 {2015},
  editor = 	 {Bach, Francis and Blei, David},
  volume = 	 {37},
  series = 	 {Proceedings of Machine Learning Research},
  address = 	 {Lille, France},
  month = 	 {07--09 Jul},
  publisher =    {PMLR},
  url = 	 {https://proceedings.mlr.press/v37/sohl-dickstein15.html}
}

@book{kay1993fundamentals,
  title={Fundamentals of Statistical Signal Processing: Estimation Theory},
  author={Kay, Steven M.},
  year={1993},
  publisher={Prentice-Hall},
  address={Englewood Cliffs, NJ, USA}
}

@article{dhariwal2021diffusion,
  title={Diffusion models beat gans on image synthesis},
  author={Dhariwal, Prafulla and Nichol, Alexander},
  journal={Advances in neural information processing systems},
  volume={34},
  pages={8780--8794},
  year={2021}
}

@misc{letafati2025diffusionmodelswirelesscommunications,
  title={Diffusion Models for Wireless Communications}, 
  author={Mehdi Letafati and Samad Ali and Matti Latva-aho},
  year={2025},
  eprint={2310.07312},
  archivePrefix={arXiv},
  primaryClass={cs.IT},
  url={https://arxiv.org/abs/2310.07312}, 
}

@INPROCEEDINGS{dmbasedchannelestimation,
    author={Ma, Xiaochuan and Xin, Yan and Ren, Yong and Burghal, Daoud and Chen, Hao and Zhang, Jianzhong Charlie},
    booktitle={2024 IEEE International Conference on Communications Workshops (ICC Workshops)}, 
    title={Diffusion Model Based Channel Estimation}, 
    year={2024},
    volume={},
    number={},
    pages={1159-1164},
    doi={10.1109/ICCWorkshops59551.2024.10615282}
}

@article{jalal2021robust,
  title={Robust compressed sensing MRI with deep generative priors},
  author={Jalal, Ajil and Arvinte, Marius and Daras, Giannis and Price, Eric and Dimakis, Alexandros G and Tamir, Jon},
  journal={Advances in neural information processing systems},
  volume={34},
  pages={14938--14954},
  year={2021}
}

@ARTICLE{10843401,
  author={Luo, Xuanhao and Li, Zhizhen and Peng, Zhiyuan and Chen, Mingzhe and Liu, Yuchen},
  journal={IEEE Transactions on Cognitive Communications and Networking}, 
  title={Denoising Diffusion Probabilistic Model for Radio Map Estimation in Generative Wireless Networks}, 
  year={2025},
  volume={11},
  number={2},
  pages={751-763},
  doi={10.1109/TCCN.2025.3529879}
}

@ARTICLE{10930691,
  author={Zhou, Xingyu and Liang, Le and Zhang, Jing and Jiang, Peiwen and Li, Yong and Jin, Shi},
  journal={IEEE Transactions on Wireless Communications}, 
  title={Generative Diffusion Models for High Dimensional Channel Estimation}, 
  year={2025},
  volume={24},
  number={7},
  pages={5840-5854},
  doi={10.1109/TWC.2025.3549592}
}

@ARTICLE{11413254,
  author={Li, Ying and Lin, Zhidi and Li, Kai and Zhang, Michael Minyi},
  journal={Journal of Communications and Networks}, 
  title={Online/offline learning to enable robust beamforming: Limited feedback meets deep generative models}, 
  year={2026},
  volume={},
  number={},
  pages={1-13},
  doi={10.23919/JCN.2025.000089}}

@ARTICLE{10937314,
  author={Zhang, Shutao and Xue, Ye and Tang, Zhiwei and Wang, Hao and Shen, Chao and Shi, Qingjiang and Chang, Tsung-Hui},
  journal={IEEE Transactions on Wireless Communications}, 
  title={Robust Network Optimization by Deep Generative Models and Stochastic Optimization}, 
  year={2025},
  volume={24},
  number={7},
  pages={6069-6084},
  doi={10.1109/TWC.2025.3551316}}

@ARTICLE{10812969,
  author={Xu, Xiaoxia and Mu, Xidong and Liu, Yuanwei and Xing, Hong and Liu, Yue and Nallanathan, Arumugam},
  journal={IEEE Communications Magazine}, 
  title={Generative Artificial Intelligence for Mobile Communications: A Diffusion Model Perspective}, 
  year={2025},
  volume={63},
  number={7},
  pages={98-105},
  doi={10.1109/MCOM.001.2400284}
}

@ARTICLE{11066175,
    author={Liu, Zhiyuan and Zhang, Shuhang and Liu, Qingyu and Zhang, Hongliang and Song, Lingyang},
    journal={IEEE Journal on Selected Areas in Communications}, 
    title={WiFi-Diffusion: Achieving Fine-Grained WiFi Radio Map Estimation With Ultra-Low Sampling Rate by Diffusion Models}, 
    year={2025},
    volume={43},
    number={11},
    pages={3796-3812},
    doi={10.1109/JSAC.2025.3584562}
}

@ARTICLE{10946972,
      author={Chen, Zhixiong and Shin, Hyundong and Nallanathan, Arumugam},
      journal={IEEE Transactions on Communications}, 
      title={Generative Diffusion Model-Based Variational Inference for MIMO Channel Estimation}, 
      year={2025},
      volume={73},
      number={10},
      pages={9254-9269},
      doi={10.1109/TCOMM.2025.3556753}
}

@misc{fan2025generativediffusionmodelswireless,
    title={Generative Diffusion Models for Wireless Networks: Fundamental, Architecture, and State-of-the-Art}, 
    author={Dayu Fan and Rui Meng and Xiaodong Xu and Yiming Liu and Guoshun Nan and Chenyuan Feng and Shujun Han and Song Gao and Bingxuan Xu and Dusit Niyato and Tony Q. S. Quek and Ping Zhang},
    year={2025},
    eprint={2507.16733},
    archivePrefix={arXiv},
    primaryClass={eess.SP},
    url={https://arxiv.org/abs/2507.16733}, 
}

@INPROCEEDINGS{10446413,
  author={Zilberstein, Nicolas and Swami, Ananthram and Segarra, Santiago},
  booktitle={ICASSP 2024 - 2024 IEEE International Conference on Acoustics, Speech and Signal Processing (ICASSP)}, 
  title={Joint Channel Estimation and Data Detection in Massive Mimo Systems Based on Diffusion Models}, 
  year={2024},
  volume={},
  number={},
  pages={13291-13295},
  doi={10.1109/ICASSP48485.2024.10446413}
}

@ARTICLE{11316498,
  author={Li, Runhua and Sun, Jian and Xue, Jiang},
  journal={IEEE Journal on Selected Areas in Communications}, 
  title={Generative Diffusion-based Bayesian Modeling for Universal Channel Estimation}, 
  year={2025},
  volume={},
  number={},
  pages={1-1},
  doi={10.1109/JSAC.2025.3648954}
}

@misc{luong2025diffusionmodelsfuturenetworks,
  title={Diffusion Models for Future Networks and Communications: A Comprehensive Survey}, 
  author={Nguyen Cong Luong and Nguyen Duc Hai and Duc Van Le and Huy T. Nguyen and Thai-Hoc Vu and Thien Huynh-The and Ruichen Zhang and Nguyen Duc Duy Anh and Dusit Niyato and Marco Di Renzo and Dong In Kim and Quoc-Viet Pham},
  year={2025},
  eprint={2508.01586},
  archivePrefix={arXiv},
  primaryClass={cs.LG},
  url={https://arxiv.org/abs/2508.01586}, 
}

@ARTICLE{10906057,
  author={Gong, Xinrui and Liu, Xiaofeng and Lu, An-An and Gao, Xiqi and Xia, Xiang-Gen and Wang, Cheng-Xiang and You, Xiaohu},
  journal={IEEE Transactions on Wireless Communications}, 
  title={Digital Twin of Channel: Diffusion Model for Sensing-Assisted Statistical Channel State Information Generation}, 
  year={2025},
  volume={24},
  number={5},
  pages={3805-3821},
  doi={10.1109/TWC.2025.3542429}
}

@INPROCEEDINGS{10888845,
  author={Bhattacharya, Sagnik and Mohsin, Muhammad Ahmed and Rajabalifardi, Kamyar and Cioffi, John M.},
  booktitle={ICASSP 2025 - 2025 IEEE International Conference on Acoustics, Speech and Signal Processing (ICASSP)}, 
  title={Successive Interference Cancellation-aided Diffusion Models for Joint Channel Estimation and Data Detection in Low Rank Channel Scenarios}, 
  year={2025},
  volume={},
  number={},
  pages={1-5},
  doi={10.1109/ICASSP49660.2025.10888845}
}

@INPROCEEDINGS{11202852,
    author={Weißer, Franz and Kasibovic, Amar and Böck, Benedikt and Utschick, Wolfgang},
    booktitle={2025 28th International Workshop on Smart Antennas (WSA)}, 
    title={No Pilots, No Problem: A Generative Model for Position-Based Downlink Precoding}, 
    year={2025},
    volume={},
    number={},
    pages={127-132},
    doi={10.1109/WSA65299.2025.11202852}
  }
\endgroup

\end{document}